\documentclass[a4paper,11pt]{article}
\pdfoutput=1
\usepackage{jheppub}
\usepackage{subfig}
\usepackage{amsmath,amssymb,bm,graphicx,epsf,colordvi}
\usepackage{slashed}
\def\bea#1\eea{\begin{align}#1\end{align}} 
\usepackage{comment}
\usepackage{diagbox}
\usepackage{mathtools}
\usepackage{lipsum}
\usepackage{soul}
\usepackage{enumitem}
\usepackage{braket}
\usepackage{bm}
\allowdisplaybreaks 
\usepackage{color}
\usepackage[dvipsnames]{xcolor}

\usepackage{mathrsfs}
\usepackage{appendix}
\usepackage{bm}
\usepackage{lipsum}

\newcommand{\bef}{\begin{figure}[hbt]\centering}
\newcommand{\eef}{\end{figure}}

\usepackage{mathtools}
\usepackage{booktabs}
\usepackage{graphicx}
\graphicspath{ {./figure/} }
\usepackage{caption}
\usepackage{lipsum}

\newcommand{\beq}{\begin{equation}}
\newcommand{\eeq}{\end{equation}}
\def\bea#1\eea{\begin{align}#1\end{align}}

\def \be  {\begin{equation}}
\def \ee  {\end{equation}}
\def \ba  {\begin{eqnarray}}
\def \ea  {\end{eqnarray}}

\allowdisplaybreaks

\usepackage{graphicx}
\usepackage{subcaption} 

\usepackage{multicol}
\usepackage[dvipsnames]{xcolor}

\usepackage{stackengine}

\title{Heavy Quark Energy Loss in the Hybrid Model}

\author[a]{Andrea Beraudo,}
\affiliation[a]{INFN --- Sezione di Torino, Via P.~Giuria 1, I-10125 Torino, Italy}

\author[b]{Jean F. Du Plessis,}
\affiliation[b]{Center for Theoretical Physics -- a Leinweber Institute, Massachusetts Institute of Technology, Cambridge, MA 02139}

\author[c,d,e]{Daniel Pablos,}
\affiliation[c]{
IGFAE, Universidade de Santiago de Compostela, E-15782 Galicia-Spain}

\affiliation[d]{Departamento de F\'isica, Universidad de Oviedo, Avda. Federico Garc\'ia Lorca 18, 33007 Oviedo, Spain}
\affiliation[e]{Instituto Universitario de Ciencias y Tecnolog\'ias Espaciales de Asturias (ICTEA), Calle de la Independencia 13, 33004 Oviedo, Spain}

\author[b]{Krishna Rajagopal}

\preprint{MIT-CTP/5950}

\abstract{
Heavy quarks offer an invaluable hard probe of the droplets of quark gluon plasma (QGP) formed in heavy ion collisions at the LHC and RHIC. Given their large mass,  they are predominantly produced in hard scattering processes at the earliest moment of a collision and given their rarity they almost never annihilate with a heavy antiquark subsequently. This means that they experience, and probe, the entire history of the expanding, cooling, droplet of QGP from hydrodynamization through hadronization. Quantitative measurements of heavy quark final state observables therefore give us access to information about the transport properties of QGP
as well as about medium modifications of hadronization. To date, the Hybrid strong/weak coupling Model of jet quenching has not included any implementation of the heavy-quark sector, which has made it impossible to confront its predictions with measurements of heavy quark and jet observables together, in a unified fashion. Here, we extend the Hybrid Model to investigate heavy quark observables for the first time. We introduce a strongly-coupled calculation of heavy-quark energy loss with the correct behavior when the heavy quarks are either ultrarelativistic or non-relativistic,
Gaussian momentum broadening, and recombination of heavy quarks with medium partons 
using a local color neutralization model of hadronization. We compare our results for the suppression $R_{\rm AA}$ and azimuthal anisotropies $v_2$ of B- and D-mesons and $\Lambda_c$ baryons, the $R_{\rm AA}$ of B-tagged jets, as well as baryon-to-meson ratios, with available experimental data from ALICE, ATLAS and CMS.
}

\begin{document}

\maketitle

\section{Introduction}
\label{intro}

Relativistic heavy ion collisions at RHIC and the LHC recreate droplets of matter with temperatures last seen throughout the universe microseconds after the Big Bang, before hadrons formed. 
Analyses of how anisotropically shaped droplets produced in 
nuclear collisions with nonzero impact parameter
explode have taught us that the matter produced in these collisions, called quark-gluon plasma (QGP), is a strongly coupled
fluid whose specific viscosity is lower than that of any other known liquid~\cite{PHENIX:2004vcz,BRAHMS:2004adc,PHOBOS:2004zne,STAR:2005gfr,Gyulassy:2004zy}. 
This has posed new challenges to understanding and opened doors to new opportunities
to study the properties and dynamics of matter governed by quantum chromodynamics (QCD), 
and indeed myriad QGP-related QCD phenomena 
have been discovered and 
analyzed with increasing incisiveness and precision over the past two decades via theoretical and experimental investigations of heavy ion collisions. 
(For reviews, see Refs.~\cite{Jacobs:2004qv,Muller:2006ee,Casalderrey-Solana:2007knd,dEnterria:2009xfs,Wiedemann:2009sh,Majumder:2010qh,Jacak:2012dx,Muller:2012zq,Mehtar-Tani:2013pia,Heinz:2013th,Shuryak:2014zxa,Akiba:2015jwa,Prino:2016cni,Romatschke:2017ejr,Connors:2017ptx,Nagle:2018nvi,Busza:2018rrf,Rapp:2018qla,Dong:2019byy,Cao:2020wlm,Schenke:2021mxx,Cunqueiro:2021wls,Apolinario:2022vzg,Harris:2023tti}.)
If the hot matter produced in a heavy ion collision were a weakly coupled plasma 
then regardless of the shapes of the droplets they would explode almost isotropically.
Instead, experimental measurements 
of the (tens of) thousands of hadrons 
produced in the explosion of droplets of QGP
at RHIC (at the LHC) 
show that the initial geometry of the overlap region between the two colliding nuclei is efficiently translated into momentum 
anisotropy --- anisotropically shaped droplets yield azimuthally anisotropic explosions.
From these experiments, we have learned that
at the temperatures that can be reached in heavy ion collisions (perhaps up to $3 T_c$,
where 
$T_c\sim 158$~MeV~\cite{Borsanyi:2010cj,Goswami:2020yez} is the temperature that characterizes the crossover from hadronic matter to QGP calculated 
using lattice gauge theory methods)
QGP is a strongly coupled fluid.
It is deconfined in the sense that it is too hot for hadrons to form, but the quarks and gluons that make up this fluid are far from free. 
Comparisons of calculations employing relativistic viscous hydrodynamics to measurements of observables that characterize azimuthal anisotropy~\cite{Teaney:2003kp,Romatschke:2007mq,Song:2007ux,Luzum:2008cw,Schenke:2010rr} (as reviewed in Refs.~\cite{Jacobs:2004qv,Heinz:2013th,Romatschke:2017ejr,Busza:2018rrf})
indicate that the specific shear viscosity $\eta/s$ (the dimensionless ratio of the shear viscosity to the entropy density) 
is lower than that of all known terrestrial liquids and is close to $1/4\pi$, its value in any large-$N_c$ gauge theory with a holographic dual in the limit of infinitely 
strong coupling~\cite{Policastro:2001yc,Policastro:2002se,Kovtun:2003wp,Buchel:2003tz,Kovtun:2004de}.
(For reviews, see Refs.~\cite{Son:2007vk,Rangamani:2009xk,Hubeny:2011hd,Casalderrey-Solana:2011dxg,DeWolfe:2013cua}.)

The droplets produced in heavy ion collisions expand and cool so rapidly that the QGP produced in this way falls apart into hadrons 
after $\sim 10^{-(22-23)}$~s, 
making it impossible to use external probes to study its properties. 
There are, however, probes that are produced in the very same collisions
in which a droplet of QGP is made. Examples include jets (collimated sprays of particles originating from a highly energetic -- typically light -- parton produced in a hard scattering that subsequently showers) and heavy quarks.
While their high energy and/or high mass guarantees that their production (and initial showering, in the case of a jet) are decoupled from the QGP dynamics, partons in a jet shower or heavy quarks will traverse the expanding cooling droplet of QGP in which they find themselves, interact with it, and exchange energy and momentum with it.  Measurements of how the observed properties of jets and heavy quarks in the final state of a heavy ion collision differ from what is seen in proton-proton collisions 
encode information about how these hard probes have interacted with QGP, meaning that they
can provide unique access to the inner workings of QGP.

The strong interactions between the high-energy partons in a jet shower and the QGP through which the shower propagates result in modifications to the parton shower and to the droplet of QGP. The resulting 
reduction to the energy and modifications to the shape and fractal structure of the jets in the final state that experimentalists measure 
are referred to in sum as 
jet 
quenching~\cite{dEnterria:2009xfs,Wiedemann:2009sh,Majumder:2010qh,Jacak:2012dx,Muller:2012zq,Mehtar-Tani:2013pia,Connors:2017ptx,Busza:2018rrf,Cao:2020wlm,Cunqueiro:2021wls,Apolinario:2022vzg}. 
This means that information about the properties and structure of QGP are encoded in the modification to the structure of a jet as it interacts with a droplet of QGP.  
Understanding how the droplet of QGP is modified by the passage of a jet through it is just as important.  The jet loses momentum to the 
QGP, exciting a wake in the expanding cooling droplet. Characterizing jet wakes, and how they propagate and dissipate, is also a way to 
learn about QGP. Because the wake that a jet creates in a droplet of QGP carries momentum in the jet direction, after hadronization these perturbations correspond to an enhancement of soft hadrons in the jet direction (which further contributes to the modification of jets as reconstructed from experimental data) as well as a depletion in the soft hadrons in the direction opposite to that of the jet.
Evidence for the latter effect has recently been reported in events with jets produced 
back-to-back with Z bosons~\cite{CMS:2025dua}.

The Hybrid Model for jet quenching~\cite{Casalderrey-Solana:2014bpa,Casalderrey-Solana:2015vaa,Casalderrey-Solana:2016jvj,Hulcher:2017cpt,Casalderrey-Solana:2018wrw,Casalderrey-Solana:2019ubu,Hulcher:2022kmn} has been developed
to analyze and understand all of the processes sketched in the previous paragraph, and more. We shall describe the model in the next Section, but in essence it starts from a conventional perturbative QCD description of jet showers in momentum space, adds a spacetime picture for how the shower develops within a droplet of QGP described via hydrodynamics, and then treats the interaction of each parton in the shower with the strongly coupled QGP wherever it finds itself 
via an energy loss formula based upon what we know about how energetic partons lose energy in a gauge theory plasma that can be described holographically at strong coupling. The Hybrid Model to date incorporates simplified treatments of the probabilistic exchange of momentum transverse to the direction of motion of a light jet parton between that parton and the QGP, as well as
the modification to the distribution of soft hadrons in the final state coming from jet wakes. 
The effects of larger-angle elastic scattering between jet partons and partons in the QGP as well as the consequences of the fact that when two jet partons are closer to each other than some QGP resolution length they lose energy to the QGP as if they were one can each 
be incorporated in the model, although we shall not consider either here.
All these strengths notwithstanding, up to now the Hybrid Model has had the weakness that it assumes that all jet partons are massless, whereas in point of fact heavy quarks are certainly found in the jets produced in RHIC- and LHC-energy collisions, and heavy quarks lose energy differently than light quarks do as they slow down. In this paper, we remedy this weakness.
Introducing heavy quarks into the Hybrid Model opens a world of new possibilities because heavy quarks are exceptionally valuable probes of the QGP in which they find themselves in their own right, whether or not they end up within a jet in the final state of a heavy ion collision.

Although a small fraction of the charm quarks in a heavy ion collision are produced via $g\rightarrow c \bar{c}$ splitting within a jet shower~\cite{Attems:2022ubu,Attems:2022otp,Brewer:2025wol}, most of them and essentially all of the bottom quarks are produced during the very initial stage of the collision, in hard processes occurring during the crossing of the two incident nuclei.  Charm and bottom quarks of course decay, but their decays occur via the weak interaction which means that they decay outside the droplet of QGP, and in fact long after the droplet of QGP has expanded and cooled.  Because the masses of both charm and bottom quarks are much larger than the temperature of the droplet of QGP even at its hottest, the possibility of thermal production of heavy quark-antiquark pairs ($Q\bar Q$) is tremendously Boltzmann suppressed and can be neglected.   
The possibility of creating $Q\bar Q$ pairs
during the process of hadronization is also completely negligible~\cite{Sjostrand:2006za}.
In light of all these features, there is a one-to-one correspondence between a heavy quark detected in the final state as a heavy meson or baryon via its weak decay at a displaced vertex and a heavy quark produced at the very beginning, at time $t=0$.  So, a charm or bottom quark seen by an experimentalist has traversed every stage of the evolution of the medium produced in a heavy ion collision, including the pre-hydrodynamic initial stages, hydrodynamization, the entire hydrodynamic evolution of an expanding cooling droplet of QGP, and hadronization at freezeout. 
The prospect of decoding information about these processes and about the properties of the QGP medium encoded in experimental measurements of heavy quark observables has motivated myriad experimental efforts at RHIC and at the LHC (for reviews, see Refs.~\cite{Prino:2016cni,Rapp:2018qla,Dong:2019byy,Apolinario:2022vzg})
and is a central motivation for the design and development of a re-envisioned ALICE3 detector for LHC Runs 5 and 6 in the 2030's and 40's~\cite{ALICE:2013nwm,ALICE:2022wwr}. 
Correspondingly, this prospect has also motivated substantial theoretical advances over several decades in the description of heavy quark transport~\cite{Moore:2004tg,vanHees:2004gq,vanHees:2005wb,vanHees:2007me,He:2011qa,Cao:2013ita,Das:2013kea,Cao:2014fna,Li:2020umn,Blok:2020jgo} as well as heavy quark recombination and hadronization~\cite{Fries:2003kq,Greco:2003vf,Greco:2004rm,Ravagli:2007xx,Fries:2008hs,Plumari:2017ntm,Beraudo:2022dpz,Beraudo:2023nlq,Zhao:2023nrz}.
For reviews, see Refs.~\cite{Prino:2016cni,Rapp:2018qla,Dong:2019byy,Apolinario:2022vzg}.

Some heavy quarks that are produced with a large initial $p_T$ lose so much momentum to the droplet of QGP that they end up almost at rest in the local fluid rest frame, diffusing in the flowing droplet of QGP.  Others that lose less momentum are found within a jet in the final state. And there are yet others that are produced with low $p_T$ initially, unrelated to any jet, that diffuse within the droplet of QGP throughout its hydrodynamic evolution.
By adding a perturbative QCD treatment of heavy quark production, a model of heavy quark energy loss that incorporates what we know from holographic calculations of non-relativistic and ultrarelativistic quarks in strongly coupled plasma, a simplified model of transverse momentum exchange between the heavy quark and the QGP,  
and the Local Color Neutralization (LCN) model
(introduced in Refs.~\cite{Beraudo:2022dpz,Beraudo:2023nlq}
and described in Section~\ref{sec:LCN})
that
describes the heavy-quark-to-heavy-hadron transition, incorporating the momentum imparted to the heavy hadron by light quark(s) drawn from the flowing QGP
into the Hybrid Model we can ultimately obtain a fully unified description of all three of these cases and everything in between.

In at least three respects, our implementation of heavy-quark energy loss in the Hybrid Model is not yet fully unified. First, there is no existing analytic holographic calculation of how a quark that is much heavier than the temperature of the QGP in which it finds itself but whose mass is not infinite loses energy, in particular if it starts out relativistic and ends up non-relativistic. For quarks with masses larger than the temperature, existing calculations are in the late time and small velocity limit. For our phenomenological purposes, we shall introduce a composite formula for heavy quark energy loss that has the correct behavior in both the ultrarelativistic and the 
non-relativistic limits and in which the energy loss $dE/dx$ changes continuously as the heavy quark loses energy. Future improvement in the holographic calculation
of heavy quark energy loss could yield a fully unified approach.
The second point where we need to introduce 
a phenomenologically motivated prescription
arises in the treatment of heavy quark hadronization.
A heavy quark can either hadronize with a soft parton from the flowing medium, obtaining momentum from the flow, or it can hadronize with partons from the parton shower
originating from the same hard scattering 
at which the heavy quark itself was produced.
The latter is well modeled by Lund string hadronization in PYTHIA, and as noted above we shall model the former via the Local Color Neutralization model~\cite{Beraudo:2022dpz,Beraudo:2023nlq}.
We shall introduce a prescription for choosing which heavy quarks hadronize in each of these two fashions, treated as distinct in our modeling.
Here also, there is room for future improvement.
Third, in our treatment of the exchange of momentum  transverse to the heavy quark direction between the heavy quark and the QGP, we shall assume that the probability distribution for the exchanged transverse momentum is Gaussian with the width of the Gaussian distribution tied to the rate of energy loss via the Einstein relation. This is a reasonable approximation for non-relativistic heavy quarks, and is the basis of 
extant treatments of heavy quark
transport via Boltzmann, Fokker-Planck, or Langevin equations.  It is however not a good approximation for relativistic heavy quarks, either for heavy quarks in weakly coupled plasma~\cite{Moore:2004tg} or in  strongly coupled plasma with a holographic dual~\cite{Gubser:2006nz,Rajagopal:2025ukd}, as has more recently been understood in any quantum field theory~\cite{Rajagopal:2025rxr}. Implementing the phenomenological implications from these recent developments into the Hybrid Model constitutes a third opportunity for further improvement along the path to a fully unified treatment of the energy loss of heavy quarks, from ultrarelativistic to non-relativistic. 

Notwithstanding the opportunities for further improvement,
we find that by implementing heavy quarks, from ultrarelativistic to nonrelativistic, losing energy in a strongly coupled fashion via the composite energy loss formula that we introduce here
into the Hybrid Model and modeling 
hadronization of a heavy quark with a light quark from the flowing QGP
using local color neutralization we 
achieve a 
rather
good description of both charm and bottom observables including the suppression ($R_{\rm AA}$) and elliptic flow ($v_2$) of
prompt and non-prompt $D^0$ mesons and $b$-tagged jets, the $p_T$-dependence of charmed baryon-to-meson ratios, as well as the $v_2$ of the $\Lambda_c$ charmed baryon.  These results from this initial study provide strong motivation for further development of the implementation of heavy quark energy loss and hadronization in the Hybrid Model.

In Section~\ref{sec:HQ-Hybrid} we introduce the Hybrid Model and describe how we shall treat the energy loss of heavy quarks therein. We also describe our implementation of heavy quark production and transport in the Hybrid Model, including a perturbative QCD treatment
of heavy quark production and a simplified model of transverse momentum exchange between the heavy quark and the QGP. In Section~\ref{sec:LCN}, we introduce the Local Color Neutralization model that we employ to describe the hadronization of heavy quarks that recombine with a light quark from the flowing droplet of QGP.
We describe our results in Section~\ref{sec:Results}, where we compare 
our calculations of $R_{\rm AA}$ and $v_2$ for charmed mesons and baryons and bottom mesons, $R_{\rm AA}$ for $b$-jets,  as well as the $p_T$-dependence of the $D^+/D^0$, $D^+_s/D^0$ and 
$\Lambda_c/D^0$ ratios to data from measurements by the ALICE, ATLAS and CMS collaborations.
We take stock and look ahead in Section~\ref{sec:Ahead}.

\section{Heavy Quarks in the Hybrid Strong/Weak Coupling Model}
\label{sec:HQ-Hybrid}

The Hybrid Strong/Weak Coupling Model (or just the Hybrid Model)~\cite{Casalderrey-Solana:2014bpa,
Casalderrey-Solana:2015vaa,
Casalderrey-Solana:2016jvj,
Hulcher:2017cpt,
Casalderrey-Solana:2018wrw,
Casalderrey-Solana:2019ubu} is a theoretical framework for describing jet formation, evolution, and quenching within a droplet of QGP produced in a heavy ion collision. 
We begin in Subsection~\ref{sec:HybridModelIntro} by introducing this model as it is in the prior literature, which is to say upon assuming that all partons in the jet showers
that it describes are massless.  In the next four subsections, we describe our treatment of the energy loss (Subsections~\ref{sec:HQ-EnergyLoss} and ~\ref{sec:Composite}) and momentum diffusion (Subsection~\ref{sec:Diffusion}) of heavy quarks and how we implement each in the Hybrid Model, as well as how we introduce realistic
charm and bottom quark spectra prior to quenching (Subsection~\ref{sec:FONLL}). 

\subsection{The Hybrid Model for Massless Quarks and Gluons in Jets in QGP}
\label{sec:HybridModelIntro}

The hard production and subsequent showering of
the energetic partons from which jets originate 
involve physics at scales that are high enough that these processes can be described using weakly coupled perturbative QCD.
In contrast,
at temperatures $T$ up to several times $T_c$, QGP 
is a strongly coupled fluid meaning that relativistic viscous hydrodynamics is the appropriate language with which to describe how a droplet of it expands and cools.
Furthermore, the 
typical interactions between
partons in the jet shower and the droplet of QGP
that result in jet quenching, modifications to the shape and structure of jets, and jet wakes in the QGP
involve momentum transfers that are of the order of the QGP temperature $T$, where QCD is strongly coupled. 
The essence of the Hybrid Model lies in describing these different regimes using the methods that are appropriate given the dominant energy scales involved in each case. 

In a heavy ion collision as in a pp collision, 
jet production begins with a hard scattering process 
characterized by a high virtuality scale $Q \sim p_T^{\rm jet} \gg T$. The high virtuality partons from such hard scattering processes 
each shower by successive splittings. As is appropriate, the Hybrid Model describes these processes using perturbative QCD, with the splittings that result in parton showers
described by DGLAP evolution equations as 
realized by PYTHIA 8~\cite{Sjostrand:2014zea}, with initial state radiation and no multi-parton interactions.
Initial state nuclear effects such as shadowing and anti-shadowing are taken into account using 
the EPS09~\cite{Eskola:2009uj} 
parametrization of parton distribution functions in the incident Pb nuclei.
In order to describe how the parton shower interacts with the droplet of QGP in which it finds itself, whose temperature depends on position and time,
the PYTHIA shower must be endowed with a spacetime structure, with a position of the initial hard scattering as well as a lifetime and velocity vector for each parton in the shower.
The point of origin of the hard scattering
in the plane transverse to the beam direction is chosen randomly with a Glauber model probability set by the density distribution of the nuclear overlap for a given centrality class.
The initial azimuthal orientation of the 
partons from the hard scattering is chosen randomly.
Each parton in the shower is then assigned a lifetime at the splitting that creates it based on its initial energy $E_i$ and virtuality $Q^2_i$ given by $\tau=2E_i/Q^2_i$~\cite{Casalderrey-Solana:2011fza}.
At each timestep, each parton in a shower is propagated with velocity
$\vec{v}=\vec{p}/E$, 
where $\vec{p}$ and $E$ are its current spatial momentum and energy, respectively.
With the spacetime structure of each parton shower generated by PYTHIA established,
these showers are then embedded into
a space- and time-dependent background that describes the expanding and cooling droplet of QGP obtained via the relativistic viscous hydrodynamics calculations of Ref.~\cite{Shen:2014vra}. 
The hydrodynamic QGP background then provides local QGP temperatures and fluid velocities at each spacetime point, necessary for computing the modifications to parton showers that are caused by their interaction with the expanding and cooling QGP droplet.

The dominant interactions between the hard partons in a jet shower and the expanding, cooling droplet of QGP involve momentum transfers at most of order $T$ and so 
are not weakly coupled processes.
So, the Hybrid Model models these interactions by assuming that an energetic quark traversing the QGP droplet loses energy at a rate with the same parametric form as that for a light quark traversing strongly coupled plasma in ${\cal N}=4$ supersymmetric Yang--Mills (SYM) theory in the limit of infinite coupling and large $N_c$. 
Although there are no known theoretical techniques for calculating light quark energy loss at strong coupling in QCD, in the ${\cal N}=4$ SYM theory one can use holography to map this calculation onto 
calculating the dynamics of a 
string falling into a $4+1$ dimensional 
black hole horizon
in an asymptotically $AdS_5\times S_5$ 
spacetime~\cite{Chesler:2014jva,Chesler:2015nqz}.
An (initially) energetic quark-antiquark pair corresponds to a small, expanding, arc of string in the bulk, far above the horizon, close to the boundary. 
The complete thermalization of the light quark corresponds to that string endpoint falling into the black hole horizon. 
The distance $x_{\rm stop}$ that the light quark travels through the strongly coupled plasma before thermalizing, its thermalization or stopping distance, 
is much larger than $1/T$ if the initial energy of the light quark is much larger than $T$. Requiring this corresponds to requiring that the string worldsheet (the initial small arc and subsequently as it evolves)
is close to null. This turns the calculation of light quark energy loss in the gauge theory plasma into a holographic calculation of (perturbations around a) congruence of null geodesics in the AdS$_5$ black hole spacetime. By analyzing such a null congruence generated by initial conditions corresponding to an energetic quark-antiquark pair, one can derive the rate of energy loss of a light quark in the strongly coupled plasma~\cite{Chesler:2014jva,Chesler:2015nqz}.

For an energetic color charge in the fundamental representation 
that has traveled through the strongly coupled fluid for a distance $x$ in the fluid rest frame, the energy loss obtained via the holographic calculation sketched above is given by~\cite{Chesler:2014jva,Chesler:2015nqz}
\begin{equation}
\label{eq:elossrate}
   \left. \frac{dE}{dx}\right|_{\rm strongly~coupled}= - \frac{4}{\pi}\, \frac{E_{\rm in}}{x_{\rm stop}}\frac{x^2}{x_{\rm stop}^2} \frac{1}{\sqrt{1-(x/x_{\rm stop})^2}} \quad ,
\end{equation}
where $E_{\rm in}$ is the initial energy of the energetic color charge before it loses any energy to the strongly coupled plasma and where
\begin{equation}
x_{\rm stop} \equiv \frac{1}{2\kappa_{\rm sc}}
\left( \frac{E_{\rm in}}{T^4}\right)^{1/3}
\label{eq:xstop}
\end{equation}
is the
aforementioned thermalization distance.
$\kappa_{\rm sc}$ is a dimensionless free parameter 
that quantifies the strength of the coupling between energetic partons and the medium. 
It depends on the t'Hooft coupling $\lambda = g^2 N_c$ as well as on the specifics of the gauge theory.
$\kappa_{\rm sc}$ can be calculated in the strongly coupled plasma of ${\cal N}=4$ SYM theory, where it is given by $1.05\lambda^{1/6}$~\cite{Chesler:2014jva,Chesler:2015nqz}. It is expected to be smaller by a factor $\sim(3-4)$ in QCD, as the stopping distance at a given temperature is expected to be longer in QCD than in ${\cal N}=4$ SYM theory. 
In the Hybrid Model, $\kappa_{\rm sc}$ is treated as a parameter to be obtained by fitting to experimental data. The fit to measurements of the suppression of jets and high-$p_T$ hadrons in Ref.~\cite{Casalderrey-Solana:2018wrw} 
yields $\kappa_{\rm sc}=0.404$, 
smaller than its value in ${\cal N}=4$ SYM theory as expected.

Additionally, note that the holographic stopping distance of a gluon is reduced by a factor of $(C_A/C_F)^{1/3} = (9/4)^{1/3}$ when compared to that of a quark~\cite{Gubser:2008as,Chesler:2014jva,Chesler:2015nqz}, where $C_A$ and $C_F$ are the Casimirs of the adjoint and fundamental representations of the color gauge group, respectively. This means that
$\kappa_{\rm gluon} = (9/4)^{1/3} \kappa_{\rm sc}$.

Summarizing, in the Hybrid Model, we take a space- and time-dependent temperature $T$ from the hydrodynamic background and apply Eq.~(\ref{eq:elossrate}) at every timestep to each parton in a jet shower generated by PYTHIA that has been situated in space and time as we have described. 
The expression~\eqref{eq:xstop} for $x_{\rm stop}$ is evaluated using the temperature $T$ of the hydrodynamic fluid at each shower parton's location in space and time as long as $T>145$~MeV, after which it as assumed no further energy loss occurs~\cite{Casalderrey-Solana:2014bpa,Casalderrey-Solana:2018wrw}.

Energy and momentum conservation requires that 
the energy and momentum lost by partons 
in the jet shower must be 
transferred to the medium.  As the medium is a strongly coupled fluid, this means that at  the same time that interactions between the jet and the fluid degrades the energy of the jet and modifies its shape and structure, the jet excites a hydrodynamic wake in the droplet of QGP.
By momentum conservation, the momentum carried by the wake of the jet in the fluid points in the direction of the jet.  Most of this momentum is carried by moving fluid behind the jet, moving in the direction of the jet.  
After freezeout, the wake of a jet becomes an excess of soft hadrons from the QGP in the jet direction and a depletion of such soft hadrons in the opposite
direction.
This has been implemented in the Hybrid Model in a simplified fashion since the work of Ref.~\cite{Casalderrey-Solana:2016jvj}.
Here, we shall focus on heavy quark observables.
In modeling jet wakes, it is reasonable to neglect the very small number of heavy quarks diffusing in the QGP and to assume that only light hadrons are produced when the droplet of QGP freezes out.  
This motivates us to assume throughout that hadrons from jet wakes do not contribute to the observables that we are interested in here, meaning that for computational efficiency we can turn the jet wakes off in the Hybrid Model.  The one unlikely possibility that we are neglecting is that a heavy quark could hadronize with a light parton from the QGP that has picked up some extra momentum from the wake of a jet. 
In our discussion of $b$-jets we shall also neglect 
the contribution of the wake, which in that case is small because we shall look at $b$-jets with the small radius $R=0.2$.

\subsection{Heavy Quark Energy Loss}
\label{sec:HQ-EnergyLoss}

Prior to this work, in the Hybrid Model all quarks have been treated as massless, meaning that (up to color factors) all quarks and gluons lose energy in the same way according to Eq.~\eqref{eq:elossrate}. In this Subsection and the next, we investigate how to extend this expression so as to describe $dE/dx$ for heavy quarks in strongly coupled plasma in a fashion that is consistent with what we know in the regimes where we know $dE/dx$ reliably, and describe how we implement the resulting composite description of heavy quark energy loss in the Hybrid Model. 

In the holographic calculation that yields the expression \eqref{eq:elossrate} for $dE/dx$, the assumption that the quark is massless corresponds to assuming that the string endpoint follows a null geodesic in the bulk as it falls toward the AdS$_5$ black hole horizon.  String endpoints live within a flavor D7 brane, meaning that they cannot fall below the bottom of this D7 brane. In the case of a light quark (quark mass $\ll T$) the bottom of the D7 brane is below the black hole horizon, meaning that the string endpoint falls into the horizon, describing thermalization: the initially energetic light quark becomes part of the strongly coupled plasma.  

Quarks with masses $M \gg T$ are described by D7 branes whose bottoms are far above the AdS$_5$ black hole horizon: string endpoints cannot fall into the horizon, and although a heavy quark can come to rest it retains its identity as a heavy quark. As described in the Introduction, a heavy quark can be traced from its production at the earliest moment of a collision through all subsequent stages of the evolution of the matter produced in the collision. The holographic analogue of this is that the endpoint of a heavy quark string cannot fall into the black hole horizon.  In the case of an infinitely heavy quark, $M\rightarrow \infty$, the bottom of the D7 brane is at the boundary of AdS$_5$, and the endpoint of the string cannot fall at all. 
The earliest holographic calculation of the energy loss of a quark propagating through strongly coupled plasma was done for the case of $M=\infty$ in Refs.~\cite{Herzog:2006gh,Gubser:2006bz}. These authors considered an infinitely massive quark that is being pulled through the strongly coupled plasma at a constant velocity $v$ by some external agent, and computed the drag force, namely the momentum transferred to the strongly coupled plasma per unit time $dp/dt$, that this external agent must exert.
(Following Ref.~\cite{Herzog:2006gh}, we shall relate this $dp/dt$ to $dE/dx$ below.)
The holographic calculation of $dp/dt$ involves calculating the shape of a string whose endpoint
moves along the AdS$_5$ boundary with velocity $v$; the string trails behind its endpoint and hangs downward into the AdS$_5$ black hole space time, coming arbitrarily close to the horizon far behind the endpoint.  Via the holographic computation of a bulk-to-boundary propagator, the shape of this string provides an analytic description of the energy and momentum in the hydrodynamic wake that the heavy quark leaves behind in the strongly coupled fluid as it is dragged through it~\cite{Chesler:2007an,Chesler:2007sv}.
The shape of the trailing string near the boundary determines the drag force, which is given by~\cite{Herzog:2006gh,Gubser:2006bz}  
\begin{equation}\label{eq:drag}
  \frac{dp}{dt}=-\eta_D p
\end{equation}
where $p=\gamma v M$ is the heavy quark momentum, with $\gamma$ the Lorentz factor, and where the drag coefficient $\eta_D$ takes the form
\begin{equation}\label{eq:dragcoef}
  \eta_D\equiv\frac{\kappa_{\rm HQ}}{2}\frac{T^2}{M}.
\end{equation}
The holographic calculation yields an explicit result for $\kappa_{\rm HQ}$ in ${\cal N}=4$ SYM theory with 't Hooft coupling $\lambda$: $\kappa_{\rm HQ}^{{\cal N}=4}=\pi\sqrt{\lambda}$. In implementing heavy quark energy loss in QCD in the Hybrid Model, we shall treat $\kappa_{\rm HQ}$ as a free parameter, to be fit to experimental data.  As for $\kappa_{\rm sc}$, we expect $\kappa_{\rm HQ}$ to be smaller in QCD  than in ${\cal N}=4$ SYM theory.

The authors of Ref.~\cite{Herzog:2006gh} argued that although the drag force \eqref{eq:drag} was first computed for a heavy quark moving at constant speed, it could also be used to describe a heavy quark moving through the strongly coupled plasma that feels no external force and therefore slows down.  In this context, following Ref.~\cite{Herzog:2006gh} it is convenient to 
convert the expression \eqref{eq:drag} for the drag force into an expression for the rate of energy loss $dE/dt$ using $E^2=p^2+M^2$ and $\frac{d}{dt}=\frac{dx}{dt}\frac{d}{dx}=\frac{p}{E}\frac{d}{dx}.$
This implies that $dp/dt=dE/dx$, meaning that
\begin{equation}\label{eq:heavyenergy-loss}
  \frac{d E}{d x}=-\eta_D\sqrt{E^2-M^2}\ ,
\end{equation}
which can be employed to describe a heavy quark which is not being pulled and which therefore loses energy and ultimately comes to rest~\cite{Herzog:2006gh}. Note that the heavy quark comes to rest in the fluid rest frame after traveling a finite distance; when we add momentum diffusion in Subsection~\ref{sec:Diffusion}, such a heavy quark will end up diffusing in the fluid, carried along by it if the fluid is flowing. 

We now have an expression for the rate of energy loss for an infinitely massive quark, \eqref{eq:heavyenergy-loss}, as well as the expression \eqref{eq:elossrate} used to describe the rate of energy loss for massless partons in the Hybrid Model. Clearly, the next question is whether the expression \eqref{eq:heavyenergy-loss} is a good approximation to $dE/dx$ 
for finite mass quarks, and if so down to how small a value of $M/T$ it can be employed.  This question was addressed via numerical calculation of finite mass corrections to Eqs.~\eqref{eq:dragcoef} and \eqref{eq:heavyenergy-loss} in Ref.~\cite{Herzog:2006gh}. These authors found very small corrections even for $M/T$ as low as $4/3$, with corrections to $\eta_D$ between 1\% and 2\% for this (far below infinite) value of $M/T$.  We shall therefore use the expression \eqref{eq:heavyenergy-loss} for $dE/dx$ for bottom and charm quarks without being concerned about $T/M$ corrections, since even at the highest temperatures reached in LHC collisions charm quarks have $M/T > 2$ and bottom quarks are of course even heavier.   

Next, we must ask up to what heavy quark velocity (up to what value of $\gamma$) the expression \eqref{eq:heavyenergy-loss} is a good approximation to $dE/dx$. We know that it must break down in the ultrarelativistic limit for light enough quarks, since there $dE/dx$ should be well approximated by Eq.~\eqref{eq:elossrate}.
In fact, various authors~\cite{Liu:2006he,Gubser:2006nz,Casalderrey-Solana:2007ahi,Chernicoff:2008sa,BitaghsirFadafan:2008adl} have shown that the calculation that yields Eq.~\eqref{eq:heavyenergy-loss} breaks down
for $\sqrt{\gamma}>M/\sqrt{\lambda}T$, which is to say it breaks down at a large enough $\gamma$ for any finite value of the quark mass $M$.
It remains an open challenge to do the holographic calculation that describes a heavy but finite-mass quark (say with $M/T\sim 2-20$ as is relevant for charm and bottom quarks in QGP) that is initially ultrarelativistic (meaning that its rate of energy loss should be close to that in Eq.~\eqref{eq:elossrate}) but that then loses energy and slows 
to the point that $\sqrt{\gamma}<M/\sqrt{\lambda}T$, 
at which point its rate of energy loss will be well approximated by Eq.~\eqref{eq:heavyenergy-loss}.  What we shall do in the next Subsection is to fit Eqs.~\eqref{eq:elossrate} and \eqref{eq:heavyenergy-loss} together into a
composite expression for the rate of energy loss of a heavy quark that behaves as it should in the two regimes where we know how it should behave.

\subsection{Composite Description of Heavy Quark Energy Loss}
\label{sec:Composite}

\begin{figure}
    \centering
    \includegraphics[width=\linewidth]{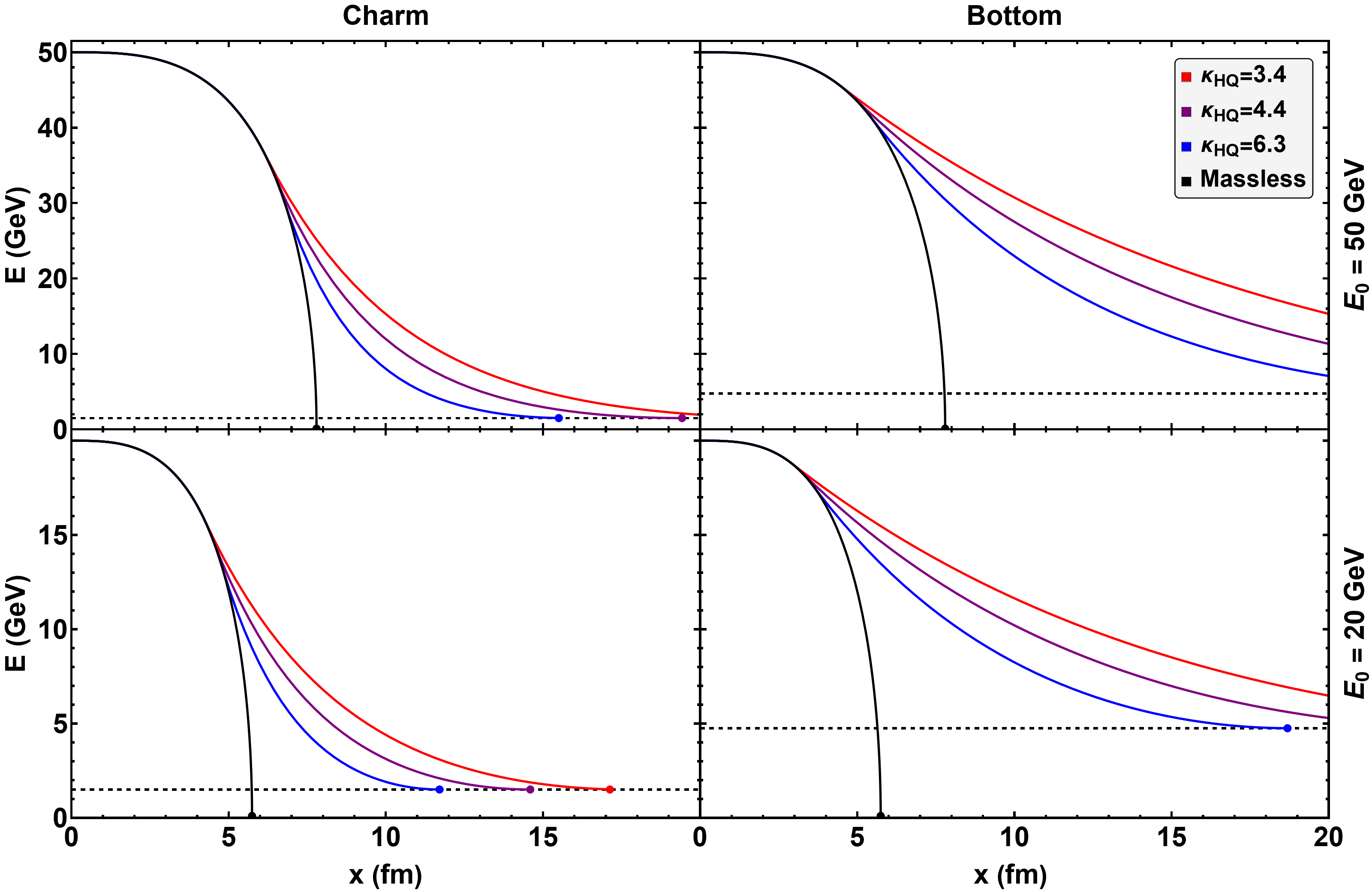}
    \caption{Comparison of how charm quarks (heavy quarks with $M=1.5$~GeV, left panels) and bottom quarks (heavy quarks with $M=4.75$~GeV, right panels) with initial energy 50~GeV (top panels) or 20 GeV (bottom panels)
    lose energy as a function of the distance $x$ that they have traveled through a ``brick'' of strongly coupled plasma with a constant temperature $T=200$ MeV. In all panels, the heavy quark initially loses energy as a light quark does. In all cases we have taken the parameter that controls the magnitude of light quark energy loss to be $\kappa_{\rm sc}=0.404$, as fit to data in Ref.~\cite{Casalderrey-Solana:2018wrw}. Curves with different colors correspond to different choices of $\kappa_{\rm HQ}$. The black curves show $dE/dx$ for a massless quark with initial energy $E_{\rm in}$, for reference. In all cases, the heavy quark comes to rest after traveling a finite distance, but in some cases this happens beyond the plot. The gray dotted line shows $E=M$, the energy of the heavy quark at rest.
    }
    \label{fig:enter-label}
\end{figure}

For our phenomenological purposes,
we 
can employ the expressions \eqref{eq:elossrate} and \eqref{eq:heavyenergy-loss} for $dE/dx$ obtained in the two limits (massless and infinitely massive) to obtain a reasonable, composite, description of a heavy quark that is initially relativistic, loses energy, and comes to rest.
It is easy to verify 
from Eq.~\eqref{eq:elossrate} and \eqref{eq:heavyenergy-loss} that the second derivative of the energy of an ultrarelativistic quark, $d^2E/dx^2$ is always negative, meaning that $dE/dx$ decreases monotonically.
And, it is easy to verify from Eq.~\eqref{eq:heavyenergy-loss}
that $d^2E/dx^2$ is always positive, with $dE/dx$ increasing monotonically.
This means that for a given $E_{\rm in}$ there is at most one value of $x$ where the two expressions for $dE/dx$ agree. 
In fact, there is always one point where this happens, since the light quark rate of energy loss  starts at 0 and diverges to $-\infty$ after it has traveled the finite distance $x_{\rm stop}$.
This leads us to propose a composite expression for $dE/dx$ for heavy quarks for our phenomenological purposes, as follows and as illustrated in Fig.~\ref{fig:enter-label}.
For a heavy quark that starts off ultrarelativistic, we expect it to lose energy like a light quark early on, and once it has lost enough energy we expect it to lose energy
as a heavy quark at late times.
So, we can simply take 
\begin{equation}\label{eq:composite}
  \frac{dE}{dx}=-\min\left(\frac{4 x^2 E_{\rm in}}{\pi x_{\rm stop}^2\sqrt{x_{\rm stop}^2-x^2}},\,\eta_D\sqrt{E^2-M^2}\right),
\end{equation}
where the quantity $x_{\rm stop}$ in the
light quark $dE/dx$ is given in terms of the parameter $\kappa_{\rm sc}$ by Eq.~\eqref{eq:xstop}
and where the drag coefficient $\eta_D$ in the 
heavy quark $dE/dx$ is given in terms of the
parameter $\kappa_{\rm HQ}$ by Eq.~\eqref{eq:dragcoef}.
Note that here $x_{\rm stop}$ is not the distance over which the heavy quark stops, since well before it stops its $dE/dx$ is specified in terms of $\eta_D$, not $x_{\rm stop}$.
The composite expression \eqref{eq:composite} for $dE/dx$
will give a continuous and  once-differentiable $E(x)$ function by construction, but note that it will have a discontinuous second derivative at the point where it switches from light to heavy quark energy loss.
For our purposes in modeling heavy quark energy loss in the QGP produced in heavy ion collisions,
both $\kappa_{\rm sc}$ and $\kappa_{\rm HQ}$
are free parameters that can be fit to data (as $\kappa_{\rm sc}$ has been in Ref.~\cite{Casalderrey-Solana:2018wrw}) or 
varied to estimate theoretical uncertainty.

In Fig.~\ref{fig:enter-label} we have plotted $E(x)$ obtained by integrating 
the composite expression \eqref{eq:composite} 
for $dE/dx$ for heavy quarks with two masses (bottom and charm quarks) with two values of the  initial energy $E_{\rm in}$
(20 and 50 GeV) in a brick of strongly coupled plasma with a constant temperature $T=200$~MeV.
We see from the Figure 
that
increasing the value of $\kappa_{\rm HQ}$ increases the rate at which the heavy quark loses energy after it begins losing energy as a heavy quark.  We also see that because of this it (slightly) delays the transition from losing energy as a light quark to losing energy as a heavy quark. 
We shall not do a formal fit to data to constrain the value of $\kappa_{\rm HQ}$ in this work, both because (as we shall see in Section~\ref{sec:Results}) the uncertainties in present experimental measurements are not yet small enough to yield a tight constraint and because this is best deferred to a future Bayesian analysis of the empirical constraints on both $\kappa_{\rm HQ}$ and $\kappa_{\rm sc}$ together. We shall find in Section~\ref{sec:Results}, though, that choosing $\kappa_{\rm HQ}=4.4$ yields a reasonable description of many quite different experimental measurements. With this in mind, in Fig.~\ref{fig:enter-label} we have chosen to plot $dE/dx$ for this value of $\kappa_{\rm HQ}$ 
as well as one higher value and one lower value.
It is reasonable to expect a value of $\kappa_{\rm HQ}$ in this vicinity for the strongly coupled QGP produced in heavy ion collisions, as $\kappa_{\rm HQ}=4.4$ is smaller by a factor of a few than in ${\cal N}=4$ SYM theory, where 
$\kappa_{\rm HQ}^{{\cal N}=4}=\pi\sqrt{\lambda}\approx 10.9$ for $\lambda=12$.

The composite expression \eqref{eq:composite} for $dE/dx$ should capture the early- and late-time behavior of a heavy quark losing energy in strongly coupled plasma well. 
We implement Eq.~\eqref{eq:composite} to describe energy loss of heavy quarks in the Hybrid Model in the same way that Eq.~\eqref{eq:elossrate} has been implemented since Ref.~\cite{Casalderrey-Solana:2014bpa}, albeit now for a quark with a velocity $v$ less than $c$.
For each heavy quark in the parton shower from PYTHIA, during each time-step $\Delta t$ in the Hybrid Model evolution,
the energy of the heavy quark is reduced by $v \Delta t \,dE/dx$, with $dE/dx$ given by the expression \eqref{eq:composite} in the fluid rest frame. We provide further details regarding the implementation of the energy loss expression~\eqref{eq:composite} in Appendix~\ref{sec:AppendixA}.
The energy and momentum lost by heavy quarks contribute to jet wakes in the same way that the energy and momentum lost by all other partons do~\cite{Casalderrey-Solana:2016jvj}, as sketched in Subsection~\ref{sec:HybridModelIntro}. In evaluating Eq.~\eqref{eq:composite}, $E_{\rm in}$ is the energy that the heavy quark had just after the most recent splitting
in the PYTHIA shower, $T$ is the local temperature of the expanding, cooling droplet of QGP described via relativistic hydrodynamics at the point in space and time where the heavy quark is found, and where, unless noted otherwise, in this study we shall fix the values of the two parameters in the model to be $\kappa_{\rm sc}=0.404$ and $\kappa_{\rm HQ}=4.4$.

After describing how we implement other important elements of heavy quark physics in the Hybrid 
Model, 
our implementation of the composite expression~\eqref{eq:composite}
will allow us to use the Hybrid Model to describe heavy quark energy loss and its observable consequences for the first time. This opens the door to a unified treatment of 
both jets and heavy quarks as probes of the droplets of QGP produced in heavy ion collisions.

\subsection{Heavy Quark Momentum Diffusion}
\label{sec:Diffusion}

As is well known and as we shall confirm via our results below, a successful description of heavy quarks with moderate and low momentum in QGP requires including the longitudinal diffusion of heavy quarks in momentum space as well as energy loss.  Transverse momentum diffusion for light quarks and gluons was implemented in the Hybrid Model in Ref.~\cite{Casalderrey-Solana:2016jvj}, and we shall employ the same 
approach here for both  longitudinal and transverse momentum diffusion, with suitable modifications for heavy quarks and upon making some simplifying assumptions. 
We shall neglect momentum diffusion when the heavy quarks are sufficiently relativistic that, per Eq.~\eqref{eq:composite}, they are still losing energy as light quarks do, as the results of Ref.~\cite{Casalderrey-Solana:2016jvj} show that the observable consequences of including momentum diffusion 
at this stage are small.  We shall see that momentum diffusion is important once the heavy quarks are less relativistic and then non-relativistic and, per Eq.~\eqref{eq:composite}, are losing energy according to Eq.~\eqref{eq:heavyenergy-loss}.
In this regime, at each time-step in the Hybrid Model evolution we give each heavy quark momentum kicks longitudinal to and transverse to its direction of motion, randomly drawn from Gaussian distributions 
with $\langle \delta p^2_L\rangle = \frac12\langle \delta p_\perp^2 \rangle = \kappa_{\rm HQ}  T^3 \gamma \Delta t = 2 T M \eta_D \gamma \Delta t$, where $\Delta t$ is the time-step in the local fluid rest frame. We have used the Einstein relation to relate $\kappa_{\rm HQ}$, defined above in terms of heavy quark drag, to heavy quark momentum diffusion, and upon doing so can now see that at low velocity $\kappa_{\rm HQ}$ is related to the conventionally defined diffusion constant $D$ and Langevin parameter $\kappa$ by $D=2/(\kappa_{\rm HQ}T\gamma)$ and $\kappa=\kappa_{\rm HQ}T^3\gamma$, which can be computed on the lattice in the zero-velocity limit.
Later in the parton shower if (as described by PYTHIA) the heavy quark splits into child partons (a heavy quark and a gluon), we reduce each child parton's initial energy at the point of the splitting by the fractional energy loss incurred by the parent parton (on account of energy loss as in  Subsection~\ref{sec:Composite} as well as momentum diffusion) and we give each child parton the same angular deflection 
incurred by the parent parton on account of momentum diffusion.

This implementation of momentum diffusion is simplified in important ways. First, we have employed the Einstein relation to relate the longitudinal momentum diffusion to the drag force, 
and it has been known since the work of Ref.~\cite{Gubser:2006nz} that although this assumption is valid for a nonrelativistic heavy quark in ${\cal N}=4$~SYM
theory it is badly violated at large $\gamma$ (by a factor of $\gamma^{3/2}$). 
We have also left out finite mass corrections to the Einstein relation (and to $\eta_D$) that are suppressed by powers of $T/M$ or $T/\gamma M$~\cite{Moore:2004tg,Beraudo:2009pe}. 
Second, we have assumed the same momentum diffusion in the directions transverse to the motion of the heavy quark as in the longitudinal direction, and this assumption is also badly violated by the same factor of $\gamma^{3/2}$ in ${\cal N}=4$~SYM
theory~\cite{Gubser:2006nz}. Third, we have assumed that
the probability distribution for the momentum picked up by the heavy quark is Gaussian, and this too is 
badly violated for relativistic heavy quarks in ${\cal N}=4$ SYM theory~\cite{Rajagopal:2025ukd}, where all the higher non-Gaussian moments of this probability distribution become important at large $\gamma$.
These facts are  related~\cite{Rajagopal:2025ukd,Rajagopal:2025rxr}: although the Einstein relation is violated, on account of the higher moments of the probability distribution heavy quarks do in fact equilibrate~\cite{Rajagopal:2025ukd}, with the Einstein relation 
replaced by a new universal condition for equilibration~\cite{Rajagopal:2025rxr}.
We leave an improved treatment of heavy quark momentum diffusion that incorporates the insights from Refs.~\cite{Rajagopal:2025ukd,Rajagopal:2025rxr}~to future work. Our goal in the present paper is to identify the 
regime of (soft) heavy hadron transverse momenta where heavy quark momentum diffusion has a substantial effect on 
suppression ($R_{\rm AA}$) and elliptic flow ($v_2$) observables and to check whether the (over)simplified treatment that we employ here suffices to give a reasonable description of extant experimental data.

\subsection{FONLL Reweighting}
\label{sec:FONLL}

\begin{figure}
    \centering
    \includegraphics[width=\linewidth]{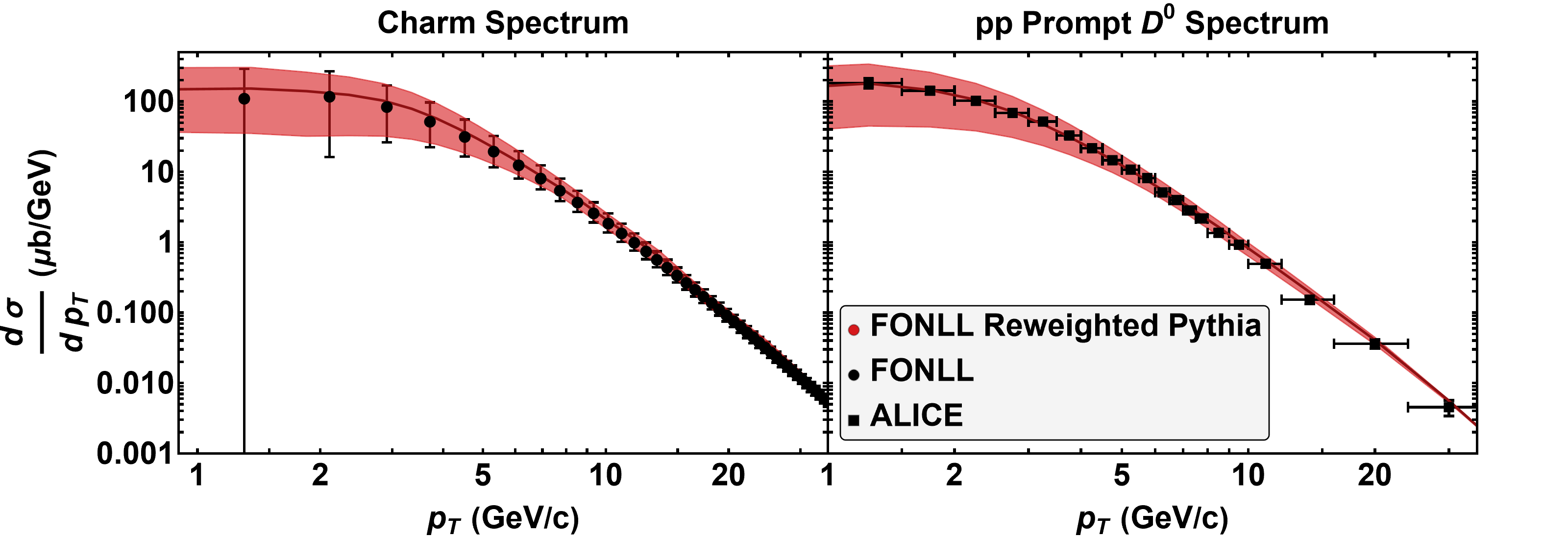}
    \caption{(Left) Charm quark spectrum from the FONLL calculation \cite{Cacciari:1998it} compared to our reweighted PYTHIA8. The reweighting is such that the error bands approximate the error bars of the FONLL calculation. This plot confirms that our reweighting works as intended. (Right) Prompt $D^0$ spectrum in pp collisions obtained from PYTHIA8, after reweighting the charm quark spectrum to agree with the FONLL calculation as in the left panel, compared to ALICE measurements from Ref.~\cite{ALICE:2021mgk}. As in the left panel, the error band quantifies the FONLL uncertainty; additional statistical errors from our Monte Carlo simulation are smaller than the thickness of the dark red line.}
    \label{plt:fonll}
\end{figure}

There is one further modification that we must make to the Hybrid Model in order to describe heavy quarks. Although PYTHIA8~\cite{Bierlich:2022pfr} provides a good description of the spectra of hadrons containing bottom quarks in pp collisions,
at low $p_T$ the PYTHIA8 charm quark spectrum fails to describe pp observables involving charmed hadrons. 
As is standard, we remedy this by reweighting 
events as a function of the $p_T$ of the highest-$p_T$ charm
quark in the event so as to to make the charm quark $p_T$ spectrum agree with that obtained from the Fixed Order Next-to-Leading Log (FONLL) calculation of Ref.~\cite{Cacciari:1998it}.
The FONLL spectrum~\cite{Cacciari:1998it} has uncertainties coming from scale and nPDF uncertainties, 
meaning that we will need slightly different reweighting schemes corresponding to the top and bottom of the error bars and to the central value from the FONLL calculation. 
We illustrate the outcome of this reweighting in Fig.~\ref{plt:fonll}, confirming in the left panel that the reweighting works as intended and showing in the right panel that it yields an excellent description of the $D^0$ spectrum measured by ALICE~\cite{ALICE:2021mgk}.
When we present the results of our calculations of 
observables in PbPb collisions in Section~\ref{sec:Results},
we shall present all observables with a darker error band representing the statistical error that results upon fixing the FONLL reweighting to its central value (the dark red lines in Fig.~\ref{plt:fonll}) as 
well as a fainter but larger error band signifying the uncertainty in the FONLL reweighting procedure,
obtained from an envelope  of the statistical error bars 
obtained upon reweighting to the top and bottom of the
FONLL error bars.  Given the excellent agreement in the right panel of Fig.~\ref{plt:fonll} between the $D^0$ spectrum obtained via reweighting the charm spectrum to the central value of the FONLL calculation with ALICE data, however, the FONLL uncertainties that we shall show in plotting our results in Section~\ref{sec:Results} (the fainter but larger error bands) are likely of auxiliary importance.

\section{Freezeout of Charmed Hadrons via Recombination}
\label{sec:LCN}

The main focus of our study is the development of a unified treatment of heavy-quark energy-loss and thermalization in a strongly coupled plasma, for heavy quarks that start out with $E_{\rm in}\gg M$ in a jet as well as for those that start out with modest initial energies.  We develop this model framework starting from the Hybrid Model which has been used to describe jets, jet shapes, jet substructure and more, and do so upon adding only a single new free parameter $\kappa_{\rm HQ}$ describing the strength of the interaction between heavy quarks and the strongly coupled plasma, a parameter that can be fit to data and that may in future be related to lattice QCD calculations of heavy quark diffusion.  This raises the prospect of studies in which we compare predictions from the model to experimental data sets coming from measurements of jet, jet substructure, and heavy quark observables together, unifying the treatment of\textemdash and lessons learned from\textemdash two of the most informative classes of hard probe observables.
This is for the future.
Our ambitious goal in the present study is to provide a satisfactory description of the experimental data 
from heavy hadron and heavy flavor jet observables
over the broadest possible kinematic range in terms of one single free parameter, leaving the light-flavor sector of the existing Hybrid Model unmodified.  

Unavoidably, experimentalists measure heavy hadrons, and jets of hadrons that include heavy hadrons\textemdash not heavy quarks themselves.   In particular for heavy hadrons that do not have a high $p_T$, a proper description of heavy quark hadronization within a hot and dense medium is mandatory in order to have a chance of describing heavy hadron observables and, subsequently, using data from measurements of these observables to constrain $\kappa_{\rm HQ}$.
In the Hybrid Model to date, with its focus on jet observables, the assumption has always been that quarks in a parton shower hadronize with other partons from the same parton shower, and hadronization has been implemented using the Lund string fragmentation module from PYTHIA.
In contrast, a heavy quark whose energy is large on account of its mass but whose momentum is not much different than the momenta of light partons from the expanding cooling droplet of strongly coupled plasma will typically hadronize by combining with a light antiquark or diquark from the medium.
Notably, the momentum that charmed hadrons acquire from the flowing medium at freezeout as they form via the above recombination process
has a significant effect on final state observables involving charmed hadrons, as we shall confirm. These effects are expected to be less significant for  (heavier) bottom hadrons. We therefore choose to focus here only on the recombination of charm quarks.

\begin{figure}[t]
    \centering
    \includegraphics[width=0.5\linewidth]{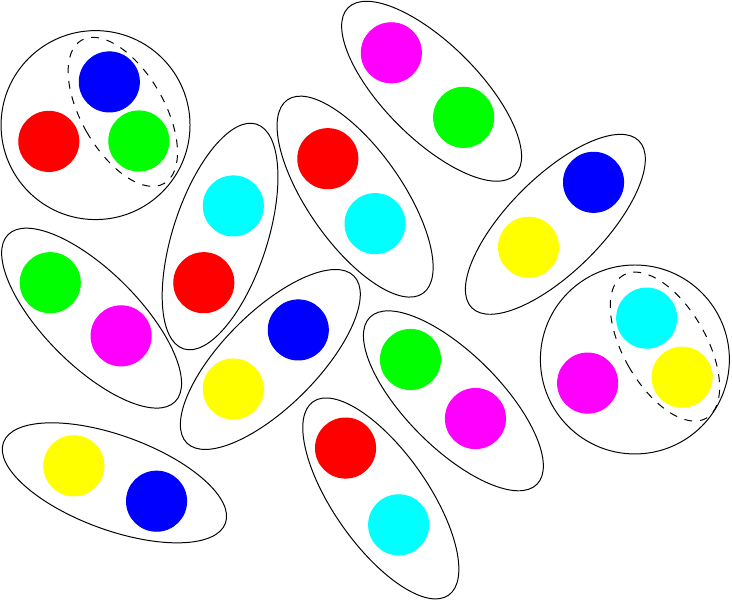}
    \caption{An illustration of the LCN hadronization model, applied in this paper to the production of charmed hadrons at freezeout. Each heavy quark recombines with the closest opposite color charge. The latter can be an antiquark, in which case recombination eventually yields a charmed meson, or it can be a diquark, in which case recombination yields a charmed baryon in the final state. In either case, as the heavy quark recombines with a light antiquark or diquark coming from the QGP the resulting color-singlet cluster picks up some momentum from the collective flow of the expanding droplet of QGP.}
    \label{fig:recom}
\end{figure}

We shall rely upon the Local Color Neutralization (LCN) model introduced in Ref.~\cite{Beraudo:2022dpz} and discussed within a broader perspective in Ref.~\cite{Beraudo:2024wgm}
to describe the formation of charmed hadrons via recombination at freezeout.
The basic idea, illustrated in Fig.~\ref{fig:recom}, is that in high-energy nuclear collisions a deconfined fireball is formed, acting as a
reservoir from which a (charm, in this case) quark can find a nearby opposite color charge to recombine
with, leading to the formation of a color-singlet cluster. 
(See Ref.~\cite{Beraudo:2022dpz} for details.)
It is reasonable to assume that recombination occurs locally, involving the nearest opposite color charge (hence the name Local Color Neutralization chosen for the model), which can be either an antiquark or a diquark. The LCN model assumes that the quark-quark attraction in the color antitriplet channel is sufficiently strong to support the existence of antitriplet diquarks as effective degrees of freedom in the QCD medium around the freezeout temperature. The presence of diquarks in the fluid cells where hadronization occurs enhances the production of charmed baryons (for similar findings within a coalescence approach see Ref.~\cite{Oh:2009zj}),
naturally explaining the enhanced baryon-to-meson
ratio observed in nuclear collisions~\cite{ALICE:2021bib}.
Note that in traditional coalescence approaches (see Ref.~\cite{Fries:2008hs} for a review) the color-singlet
cluster resulting from a $2\to 1$ or $3\to 1$ recombination process is directly identified with a physical hadron or resonance, whereas in the LCN model it should be regarded as an intermediate cluster state that subsequently decays into a charmed meson or baryon.

\begin{table}[t]
   \begin{center}  
  \begin{tabular}{|c|c|c|c|c|}
\hline
Species & $g_s$ & $g_I$ & $M$ (GeV) & daughter (if $M_{\cal C}\!<\!M_{\rm cut}$)\\
\hline
$l$ & 2 & 2 & 0.33000 & $D^0,D^+$\\
\hline
$s$ & 2 & 1 & 0.50000 & $D_s^+$\\
\hline
$(ud)_0$ & 1 & 1 & 0.57933 & $\Lambda_c^+$\\
\hline
$(ll)_1$ & 3 & 3 & 0.77133 & $\Lambda_c^+$\\
\hline
$(sl)_0$ & 1 & 2 & 0.80473 & $\Xi_c^0,\Xi_c^+$\\
\hline
$(sl)_1$ & 3 & 2 & 0.92953 & $\Xi_c^0,\Xi_c^+$\\
\hline
$(ss)_1$ & 3 & 1 & 1.09361 & $\Omega_c^0,\Xi_c^+$\\
\hline
   \end{tabular}
       \end{center}
\caption{The light thermal species (quarks or diquarks) involved in the recombination process, with their spin and isospin degeneracy, their mass and the daughter charmed hadrons arising from the cluster decay. In the above, $l$ can be either $u$ or $d$, assuming isospin symmetry.  In the LCN model, when a charm quark recombines with the species in the left column, it forms a cluster ${\cal C}$. As long as the invariant mass of the cluster is less than $M_{\rm cut}$ (a parameter that we set to 3.8 GeV) it decays into one of the daughter charmed hadrons in the right column, as well as a light hadron or photon. Heavier clusters hadronize via fragmentation, not via recombination.}
\label{tab:masses}
\end{table}

We now provide a simplified description of the implementation of the LCN model adopted in this paper, referring the reader to Ref~\cite{Beraudo:2022dpz} for further details.
Once a charm quark, whose propagation in the droplet of strongly coupled plasma is described by the Hybrid Model
via the composite energy loss expression~\eqref{eq:composite}
and the stochastic momentum diffusion described in 
Section~\ref{sec:Diffusion},
reaches a fluid-cell on the freezeout hypersurface --- where the fluid temperature is 
$T_{\rm freezeout}=145$ MeV (which is the temperature at which parton energy loss is assumed to stop in the Hybrid Model~\cite{Casalderrey-Solana:2014bpa,Casalderrey-Solana:2018wrw}, and which differs from the value 155 MeV employed in Ref.~\cite{Beraudo:2022dpz}) --- it is recombined with a thermal particle from the same cell. The latter can be either a light antiquark ($u,d,s$) or a diquark.
For their masses, summarized in Table~\ref{tab:masses}, we take the default values employed by PYTHIA 6.4~\cite{Sjostrand:2006za}. Light quarks and diquarks are assumed to be distributed in chemical and kinetic equilibrium in the local rest frame of the fluid cell. In this frame, the species and momentum of the charm-quark companion can then be sampled from a thermal distribution and a color-neutral cluster ${\cal C}$ is then formed. As is done in the Herwig event generator~\cite{Webber:1983if}, an intermediate cutoff $M_{\rm cut}$ on the invariant mass of the cluster $M_{\cal C}$ is introduced, meaning that if the  cluster is sufficiently light ($M_{\cal C}<M_{\rm cut}$) it is assumed to undergo an isotropic two-body decay (into a charmed hadron and a light hadron\textemdash usually a pion\textemdash or a photon) in its own rest frame. The possible daughter charmed hadrons are listed in Table~\ref{tab:masses}. 
On the other hand, if at least one  cluster
has a larger invariant mass ($M_{\cal C}>M_{\rm cut}$), 
the two heavy quarks coming from this hard scattering process are each
treated as color connected with the rest of the parton shower in which they formed, and we simulate their hadronization using the PYTHIA string-fragmentation module, as is done for massless partons in a jet shower in the Hybrid Model. Heavy hadrons do not undergo rescattering after freezout in either mode of hadronization in our model. We similarly hadronize any charmed quarks in a Hybrid Model shower that were either created by a gluon splitting outside the freezeout hypersurface or that radiated a gluon outside the freezeout surface via string-fragmentation as usually done in the Hybrid Model. 
In the following, we set  $M_{\rm cut}=3.8$ GeV, as in 
Ref.~\cite{Beraudo:2022dpz}. 
(In the Herwig event generator~\cite{Webber:1983if}, $M_{\rm cut}=4$ GeV is the maximum invariant-mass of clusters undergoing a two-body decay.) We have checked that varying the value of $M_{\rm cut}$ over the range 3-5~GeV makes very little difference to any of the observables that we have calculated. Both $M_{\rm cut}$ and $T_{\rm freezeout}$ are parameters in the model; in a future more systematic study one could vary both over reasonable ranges to estimate the associated (small) contributions to the systematic uncertainty in results obtained from the model.
In Section~\ref{sec:Results}, we shall focus our attention on the parameter $\kappa_{\rm HQ}$ which governs the strength of the interaction between the heavy quarks and the strongly coupled plasma. This parameter is important in that our model predictions for some heavy hadron observables are sensitive to its value, meaning that we can use experimental data to learn about its value.

\begin{figure}[t]
    \centering
    \includegraphics[width=\linewidth]{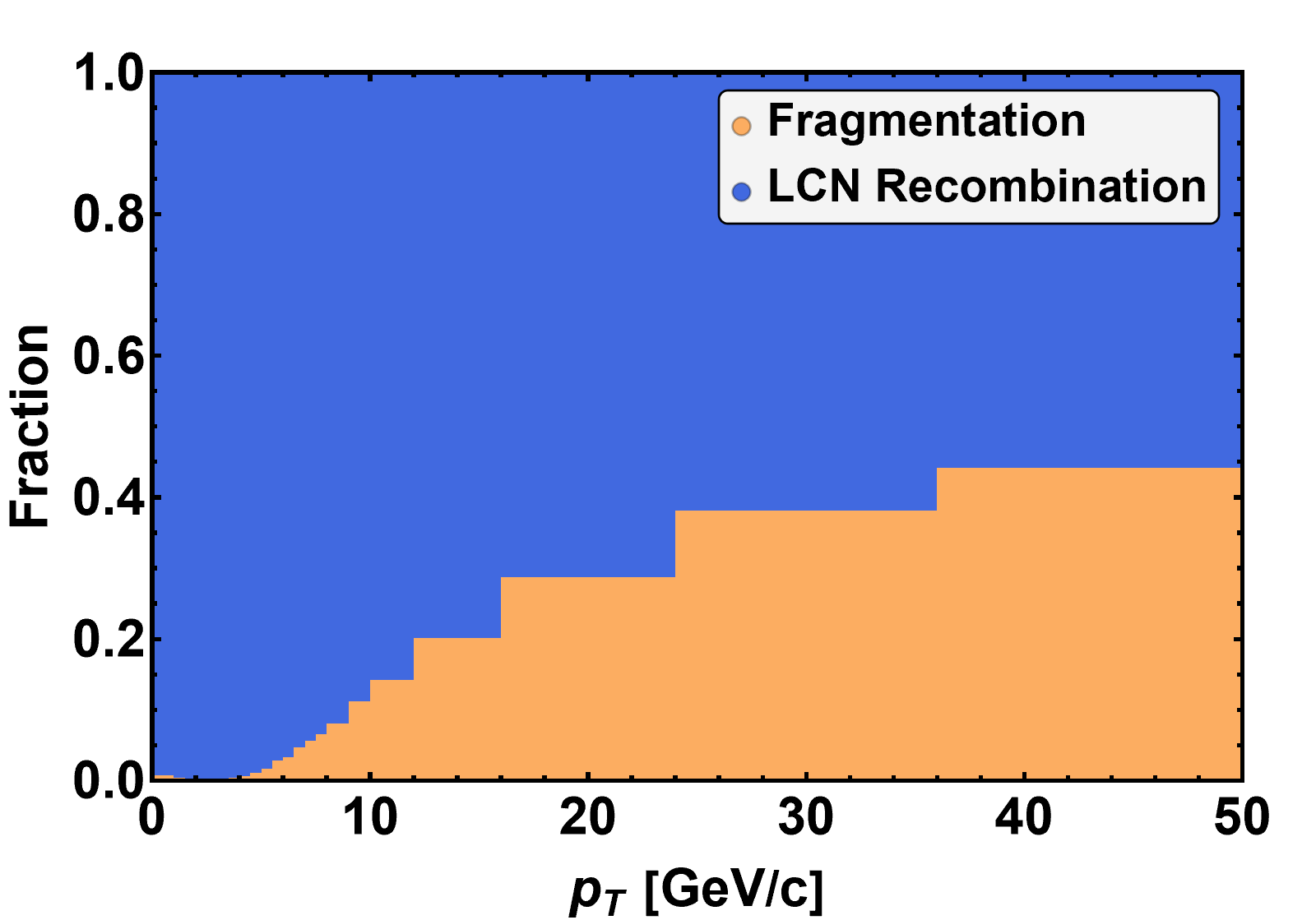}
    \caption{Fraction of $D^0$ mesons with a given $p_T$ in the 0-5\% centrality class that formed either via LCN recombination or via fragmentation as described by PYTHIA 8. Almost every $D^0$ meson with $p_T$ less than about twice its mass (which is to say most $D^0$ mesons) form via recombination. At these low values of $p_T$ (and also at intermediate $p_T$), the momentum that the charm quark picks up from the flowing droplet of QGP via recombination with a light antiquark contributes significantly to the momentum and, we shall see, to the $v_2$ of the $D^0$ mesons.}
    \label{fig:hadform}
\end{figure}

With this choice of parameters, the condition $M_{\cal C}<M_{\rm cut}$ is indeed satisfied in almost every case for 
charm quarks with $p_T$ of order or less than the charm quark mass, 
meaning that most charmed 
hadrons form via LCN recombination.
At the same time, as the transverse momentum of the parent charm quark increases, the fraction of clusters for which $M_{\cal C}>M_{\rm cut}$ also grows, so that our model allows a smooth interpolation from the regime of low-to-moderate $p_T$ -- where charmed-hadron production occurs via recombination -- to the high-$p_T$ tail of the momentum distribution, where string fragmentation plays a significant role, as illustrated in Fig.~\ref{fig:hadform}.  LCN remains relevant for high-$p_T$ charmed hadrons, though, because a high $p_T$ charm quark can recombine with a thermal antiquark or diquark from the QGP that is close to it in position space as long as the angle  of the diquark or antiquark in momentum space is close enough to that of the charm quark.
In this case, the ultrarelativistic charm quark and the light antiquark or diquark have similar velocities and stay close to each, meaning that they can form a low-invariant mass cluster via {\it local} color neutralization.  We see in Fig.~\ref{fig:hadform} that $D$ mesons with $p_T\sim 50$~GeV can be formed either via fragmentation or via LCN recombination, with comparable probability.

The local nature of the recombination process within an expanding droplet of QGP
means that a heavy quark and the light antiquark or diquark with which it recombines in the LCN model are close to each other in angle as well as in position space. 
This gives rise to the formation of low invariant-mass clusters,
most often decaying into just two final-state particles (the charmed hadron and a light hadron or photon) as we have described above.  Furthermore, the charmed hadron inherits
some of its momentum from the collective flow of the 
expanding droplet of QGP.
In the next Section, we shall see
the implications of charmed hadron formation via recombination for the relative yields of various charmed hadrons as well as for their spectra and elliptic flow, in particular at modest $p_T$ where the contribution to their momentum originating from the collective flow of the droplet of QGP is important. By turning LCN recombination off and on we shall see that including it is crucial to describing charmed hadron observables at modest transverse momenta.

\section{Comparing Hybrid Model Calculations of Heavy Hadron Observables to Experimental Data}
\label{sec:Results}

With our composite description of heavy quark energy loss, FONLL initial conditions for charm quarks, heavy quark momentum diffusion, and LCN recombination of charm quarks all implemented in the Hybrid Model,
we have laid all the groundwork needed to compute heavy hadron observables in the Hybrid Model. Furthermore, we can do so both for high-$p_T$ heavy flavor hadrons that are found within jets and for low-$p_T$ heavy flavor hadrons that need not have any association with a jet.  In this Section, we compare Hybrid Model calculations of a suite of heavy hadron observables to experimental measurements.  We shall set $\kappa_{\rm sc}=0.404$ throughout~\cite{Casalderrey-Solana:2018wrw}, and through much of this Section we shall show results obtained from the Hybrid Model with $\kappa_{\rm HQ}=4.4$. This choice is discussed in Section~\ref{sec:kappaval}.

\subsection{$R_{\rm AA}$ and $v_2$ for Charm and Bottom Mesons}
\label{sec:mainresults}

\begin{figure}[t]
    \centering
    \includegraphics[width=\linewidth]{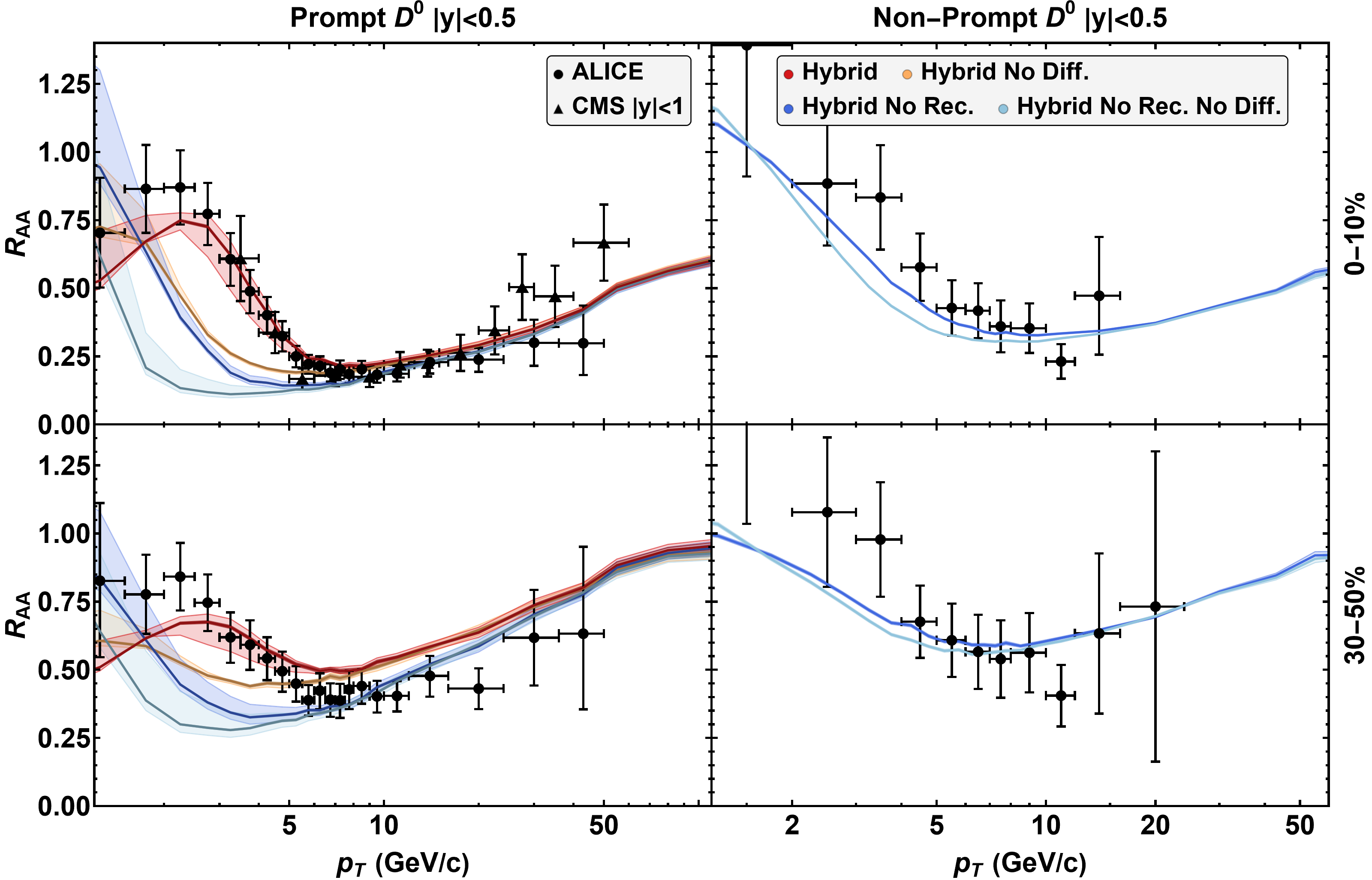}
    \caption{Results from Hybrid Model calculations with $\kappa_{\rm HQ}=4.4$ of the suppression $R_{\rm AA}$ of $D^0$ mesons in PbPb collisions at $\sqrt{s_{\rm NN}}=5.02$~TeV compared to ALICE \cite{ALICE:2021rxa,ALICE:2021mgk} and CMS \cite{CMS:2017qjw} data. Hybrid Model calculations and ALICE data use a rapidity cut of $|y|<0.5$, while the CMS data is for $|y|<1$. The left (right) panels show the suppression of prompt (non-prompt) $D^0$ mesons, originating from $c$-quarks ($b$-quarks) propagating in the droplet of QGP. 
    The upper (lower) panels are for 0-10\% (30-50\%) centrality collisions. For each experimental data point, we have added the statistical and systematic uncertainties in quadrature. For each Hybrid Model calculation, the darker (fainter) band shows the statistical uncertainties (uncertainties coming from FONLL reweighting of the $c$-quark production spectra).} 
    \label{fig:hadraa}
\end{figure}

\begin{figure}[t]
    \centering
    \includegraphics[width=\linewidth]{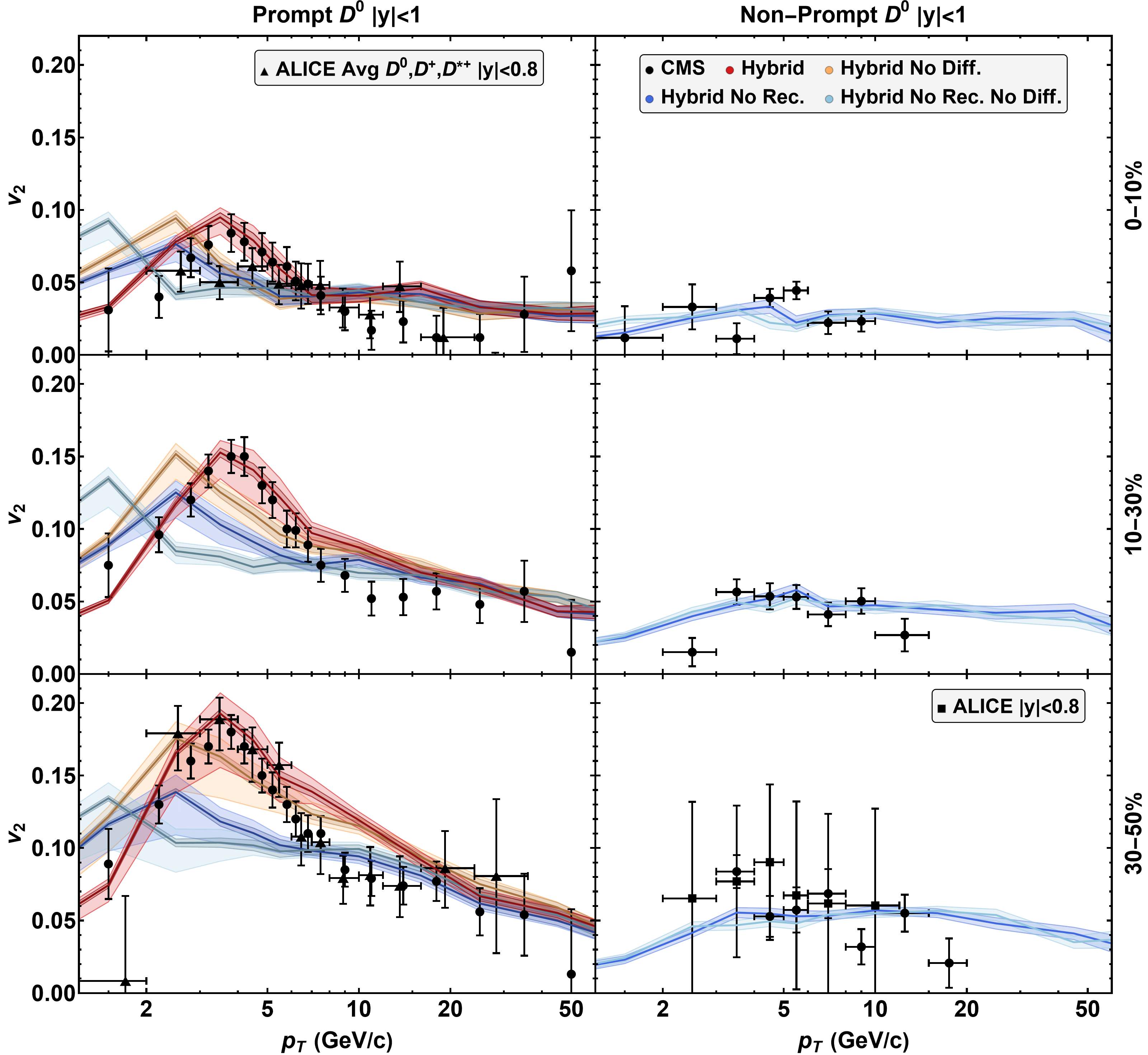}
    \caption{Results from Hybrid Model calculations with $\kappa_{\rm HQ}=4.4$ of the azimuthally anisotropic elliptic flow $v_2$ of $D^0$ mesons in PbPb collisions at $\sqrt{s_{\rm NN}}=5.02$~TeV compared to ALICE \cite{ALICE:2020iug,ALICE:2023gjj} and CMS \cite{CMS:2020bnz,CMS:2022vfn} data.  Hybrid Model calculations and CMS data use a rapidity cut of $|y|<1$, while the ALICE data is for $|y|<0.8$. The left (right) panels shows the suppression of prompt (non-prompt) $D^0$ mesons, originating from $c$-quarks ($b$-quarks) propagating in the droplet of QGP. 
    The upper, middle and lower panels are for 0-10\%, 10-30\% and 30-50\% centrality collisions. For each experimental data point, we have added the statistical and systematic uncertainties in quadrature. For each Hybrid Model calculation, the darker (fainter) band shows the statistical uncertainties (uncertainties coming from FONLL reweighting of the $c$-quark production spectra).}
    \label{fig:hadv2}
\end{figure}

We begin in Figs.~\ref{fig:hadraa} and \ref{fig:hadv2} by confronting Hybrid Model calculations of the suppression ($R_{\rm AA}$) and elliptic flow ($v_2$) of both charmed mesons (specifically, prompt $D^0$ mesons) and bottom hadrons (specifically, those detected via the measurement of non-prompt $D^0$ mesons) to experimental measurements of these key observables by ALICE and CMS. 
In our calculations, we compute $v_2$ via the expression $v_2=\langle(p_x^2-p_y^2)/(p_x^2+p_y^2)\rangle$. This simple expression is appropriate since we are using a single event-averaged hydro profile whose impact parameter points in the $x$-direction for each centrality bin. 
In our Hybrid Model calculations, we record
all
$D^+,D^0,D_s,\eta_c$ and $J/\psi$ mesons as well as all charmed baryons. We can then check whether each charmed meson or baryon --- for example each $D^0$ meson --- came from the weak decay of a $b$-meson or $b$-baryon long after and far outside the droplet of QGP. 
Those that did not, namely prompt $D^0$ mesons, give us access to the physics of $c$-quark propagation in QGP. 
Non-prompt $D^0$ mesons 
give us access to the physics of $b$-quark propagation in QGP.  
We note that CMS has measured $R_{\rm AA}$  for $B^+$, $B_s^0$ and $B_c^+$ mesons themselves in PbPb collisions with 0-90\% centrality~\cite{CMS:2022sxl,CMS:2024vip}; we defer the analysis of these and other (almost) minimum-bias observables to future work.  

Before we begin discussing the results in Figs.~\ref{fig:hadraa} and \ref{fig:hadv2}, it is useful to clarify the meaning of the different 
Hybrid Model calculations shown as bands with different colors in these and subsequent Figures.
In the left panels showing prompt $D^0$ meson observables, the red bands (labeled as hybrid) show results obtained from full Hybrid Model calculations, including momentum diffusion and recombination. 
 For the purpose of understanding the relative contribution of those different effects across hadron $p_T$ and centrality, we also include results from calculations that lack either
 recombination (Hybrid No Rec., in dark blue) or momentum diffusion (Hybrid No Diff., in orange), or from calculations with both momentum diffusion and recombination turned off  (Hybrid No Rec. No Diff., in light blue). We will see that this way of dissecting our results, by switching certain physical effects on or off in a way that we can do in a model but that is impossible to do in data, offers key insights into the interaction between the heavy quarks and the flowing droplet of QGP and their approach to thermalization. In the right panels showing 
 non-prompt $D^0$ meson observables, our full results corresponds to the dark blue Hybrid No Rec.~curves since, as explained above, we have not included recombination between bottom quarks and light quarks from the QGP because it yields a much smaller effect on $b$-hadron observables 
 on account of the larger mass of the bottom quark. 

We now discuss the $R_{\rm AA}$ results shown in Fig.~\ref{fig:hadraa} and the $v_2$ results shown in Fig.~\ref{fig:hadv2}. 
The first observation that catches the eye is the remarkable agreement between our full results (red bands for prompt $D^0$ mesons; dark blue bands for non-prompt $D^0$ mesons) and experimental data from the ALICE \cite{ALICE:2021rxa,ALICE:2021mgk} and CMS \cite{CMS:2017qjw} collaborations for $R_{\rm AA}$ in Fig.~\ref{fig:hadraa}, and from the ALICE~\cite{ALICE:2020iug,ALICE:2023gjj} and CMS~\cite{CMS:2020bnz,CMS:2022vfn} collaborations for $v_2$ in Fig.~\ref{fig:hadv2} -- data for both hadron species, three centrality classes, and across essentially two orders in magnitude in hadron $p_T$, from 1 to almost 100 GeV. 
We also see immediately that, in particular for prompt $D^0$ mesons, we only obtain results that are in agreement with the experimental measurements if we include both charm quark momentum diffusion and charm quark recombination in our calculations.  This means that we should be able to see
how these different physical effects contribute to both $R_{\rm AA}$ and $v_2$, and it tells us that the quantitative success of the full calculation
is rather non-trivial as, especially at low $p_T$, it originates from a delicate conjunction of phenomena.

Let us frame our discussion in partonic terms first.
The hydrodynamic picture of the dynamics of a droplet of QGP states that the size of $v_2$ for low $p_T$ particles reflects the azimuthal anisotropy of the transverse flow of the droplet of fluid.
This anisotropic flow develops due to the presence of anisotropic pressure gradients caused by the anisotropic initial geometry in any heavy ion collision with a nonzero impact parameter. 
The anisotropic flow $v_2$ increases with decreasing centrality, as with increasing impact parameter the eccentricity of the initial shape of the collision zone increases.
If $\kappa_{\rm HQ}$ is very large, heavy quarks moving through the fluid feel a large drag force and rapidly lose energy to the fluid and (in the absence of momentum diffusion) would come to rest in the local fluid rest frame, and from then on would be carried along by the moving fluid.  Including momentum diffusion means that heavy quarks that have been fully quenched in this fashion end up diffusing while carried along by the moving fluid.  Assuming that the momentum diffusion is related to the drag by the Einstein relation, fully quenched heavy quarks will end up with a momentum distribution that is a Boltzmann distribution, boosted by the local fluid velocity. If this (simple) picture is correct for heavy quarks in some kinematic regime, the $v_2$ of the corresponding heavy mesons will increase with increasing impact parameter.
For an initial assessment of where this picture of heavy quark transport in QGP might be relevant, we can inspect our Hybrid Model results obtained without recombination (as recombination reflects the dynamics of hadronization, not heavy quark transport).   
If we look at the values of $v_2$ for the light blue bands in Fig.~\ref{fig:hadv2}
at, say, $p_T=3$~GeV we see that for prompt $D^0$ mesons (proxies for charm quarks) $v_2$ increases from 0.04 to 0.08 to 0.10 as the centrality increases from
0-10\% to 10-30\% to 30-50\%.  In contrast, the $v_2$ for non-prompt $D^0$ mesons (proxies for bottom quarks) only changes from 0.03 to 0.04 to 0.04.  This is strong evidence that bottom quarks  {\it do not} thermalize in, and end up flowing along with, the fluid. And, this poses the question of what is the degree to which, or the $p_T$-range over which, charm quarks {\it do} thermalize within the fluid. We shall address this in the next Subsection.

We continue our discussion in partonic terms of what we see in Figs.~\ref{fig:hadraa} and \ref{fig:hadv2} by comparing the light blue and dark blue curves so as to assess the consequences of momentum diffusion for $R_{\rm AA}$ and $v_2$.
We see that including momentum diffusion pushes both $R_{\rm AA}$ and $v_2$ to the right, does so more for charm quarks than for bottom quarks, and does so most at low $p_T$. This is readily understood since although momentum diffusion can increase or decrease the momenta of a heavy quark it is the upward fluctuations that matter when the heavy quark spectrum is a rapidly decreasing function of $p_T$. 

We turn now to assessing the effects of recombination, which is only included for charm quarks.  We can do so by
comparing the red curves in the left panels of Figs.~\ref{fig:hadraa} and \ref{fig:hadv2} to the dark blue curves, or (if we wish to avoid the confounding effects of momentum diffusion) by comparing the orange curves to the light blue curves.
In either case, what we see is that when a charm quark hadronizes via recombining with a nearby light antiquark, it picks up $p_T$ from the flowing medium, which tends to shift $p_T$ to the right in both Figs.~\ref{fig:hadraa} and \ref{fig:hadv2}.
While the size of this shift is comparable to the one obtained just from momentum diffusion (dark blue curve relative to light blue curve) in the 0-10\% centrality class shown in the top left panel of Fig.~\ref{fig:hadraa}, one can see that in the 30-50\% centrality class the effects of
recombination are  larger and extend to higher $p_T$.
We also see in the left panels of Fig.~\ref{fig:hadraa} that it is only via the interplay of the effects of momentum diffusion and recombination, together, on the prompt $D^0$ spectra that we are able to obtain a good description of the measured $R_{\rm AA}$.

That said, it is apparent from the left panels of Fig.~\ref{fig:hadv2} that recombination has more nuanced effects on the $v_2$ of prompt $D^0$ mesons
than on their $R_{\rm AA}$.
We not only observe a general recombination-induced  $p_T$-shift to the right for all centralities studied, but 
also a centrality-dependent modification of the
magnitude of $v_2$ at the $p_T$ where $v_2$
has its maximum. This trend is most neatly appreciated by comparing the light blue (neither momentum diffusion nor recombination) versus the orange (just recombination) curves. 
As the size of the elliptic flow of the bulk of the QGP fluid increases with decreasing centrality, so does the value of the elliptic flow inherited by the recombined prompt 
$D^0$ mesons. 
Without recombination, a flowing charm quark hadronizes with a light antiquark from the parton shower. In a central collision, where bulk $v_2$ is relatively low, replacing that light shower parton with a flowing parton from the medium to build up a cluster via recombination has a relatively smaller effect than in a semi-central collision where the bulk $v_2$ is notably larger.

Lastly, towards high energies, above $p_T>30$ GeV, recombination effects also become irrelevant, as one would expect. Moreover, at these high values of $\gamma$  the heavy quark is behaving as a light quark, and generally losing energy according to the light parton energy loss rate as prescribed by Eq.~\eqref{eq:composite}. Therefore, the satisfactory agreement between Hybrid Model results and experimental data observed for the $R_{\rm AA}$ of heavy mesons at high $p_T$ in Fig.~\ref{fig:hadraa} is heavily tied to the satisfactory agreement between Hybrid Model calculations and experimental measurements of the suppression of light charged hadrons and inclusive jets obtained
in the previous work~\cite{Casalderrey-Solana:2018wrw}, 
where the value of $\kappa_{sc}=0.404$ was fitted. 
The very good agreement between Hybrid Model calculations and experimental measurements of high-$p_T$ $v_2$ observed in Fig.~\ref{fig:hadv2} is nonetheless noteworthy, as only $R_{\rm AA}$ was fitted in Ref.~\cite{Casalderrey-Solana:2018wrw}.
It has long been understood~\cite{Gyulassy:2000gk,Wang:2000fq}
that the azimuthal anisotropy $v_2$ of high-$p_T$ hadrons originates from an interplay between energy loss and selection bias. In heavy ion collisions with a nonzero impact parameter, high-$p_T$ partons traveling along the event plane will on average propagate through less QGP and lose less energy than those traveling perpendicular to the event plane. 
This, together with the fact that the  parton spectrum at high $p_T$ is steeply falling, means that partons propagating along the event plane, and the hadrons that result, represent a larger relative percentage of the population of hadrons with any fixed large
$p_T$. This argument applies just as much to heavy partons, and the heavy hadrons that result, as to light partons.

Summing up, we conclude that the Hybrid Model yields a good qualitative and reasonably quantitative description of the $R_{\rm AA}$ and $v_2$ of prompt $D^0$ mesons, proxies for charm quarks, 
only when both momentum diffusion and recombination effects are included. 
While momentum diffusion does improve the Hybrid Model description of the $R_{\rm AA}$ for non-prompt $D^0$ mesons somewhat,
both because of the smallness of the effects and the relatively larger size of the experimental error bars, no strong conclusion can be drawn. 
Together with the excellent agreement between Hybrid Model calculations and ATLAS measurements of jet $v_2$~\cite{ATLAS:2013ssy} found in previous work~\cite{Du:2021pqa} and together with our own comparison between Hybrid Model calculations of $b$-jet $R_{\rm AA}$ with experimental measurements in Subsection~\ref{sec:jetRAA},
the new Hybrid Model results for the
$v_2$ and $R_{\rm AA}$ of heavy hadrons that we have described above
demonstrate the ability of the Hybrid Model to 
provide a natural, seamless, simultaneous
description of $R_{\rm AA}$ and $v_2$ for light hadrons, heavy hadrons and jets.

There are of course also indications of the need for improvements in our framework. The curves excluding recombination effects (light and dark blue) describe 30-50\% prompt $D^0$ meson $R_{\rm AA}$ data (bottom row of Fig.~\ref{fig:hadraa})  and both 10-30\% and 30-50\% $v_2$ data (middle and bottom rows of Fig.~\ref{fig:hadv2}) somewhat better than the ones which do include them (orange and red) for intermediate $p_T$ values, namely $8\lesssim p_T\lesssim 20$ GeV.
This observation suggests that our switchover criterion between hadronization via fragmentation and hadronization via LCN recombination described in Section~\ref{sec:LCN}
may be too crude and deserves a more careful treatment, which we defer to future work. 
The reader can also (correctly) surmise that increasing the value of $\kappa_{\rm HQ}$ would improve the agreement between the $R_{\rm AA}$ curve displaying our full results (in red, with recombination) and experimental data. However, doing so would decrease agreement between our full results for $v_2$ and experimental data in that range. The nature of this tradeoff and a brief discussion of the (exploratory) choice of the value of $\kappa_{\rm HQ}=4.4$ is addressed in Subsection~\ref{sec:kappaval}.

\subsection{The Degree of Thermalization of Heavy Quarks}

By comparing the lack of centrality dependence in the right panels of Fig.~\ref{fig:hadv2} to the centrality dependence 
seen in the left panels, we have seen in the previous Subsection that bottom quarks do not thermalize in the flowing QGP whereas charm quarks do so to some degree.
The relative lack of thermalization of bottom quarks relative to charm quarks arises from two effects.
First, as we have seen in Fig.~\ref{fig:enter-label}, the stopping length for bottom quarks is longer than for charm quarks with the same initial energy because the drag coefficent $\eta_D$ is proportional to $T/M$. This makes it less likely that bottom quarks can thermalize in the finite droplets of QGP produced in a heavy ion collision. 
Second, if we imagine comparing a bottom quark and a charm quark that initially have zero velocity that
find themselves in the QGP fluid with some nonzero transverse velocity, the bottom quark needs to pick up more transverse momentum (via the combination of drag and momentum diffusion) in order to equilibrate with the moving fluid.
Furthermore, a recent measurement 
made by CMS~\cite{CMS:2025hki} 
shows that in each centrality class there exists an event-by-event linear correlation between the $v_2$ of the bulk event and that of the prompt $D^0$ mesons in the event. This, too, suggests that charm quarks must, to some degree, end up flowing along with the QGP fluid. 

\begin{figure}
    \centering
    \includegraphics[width=\linewidth]{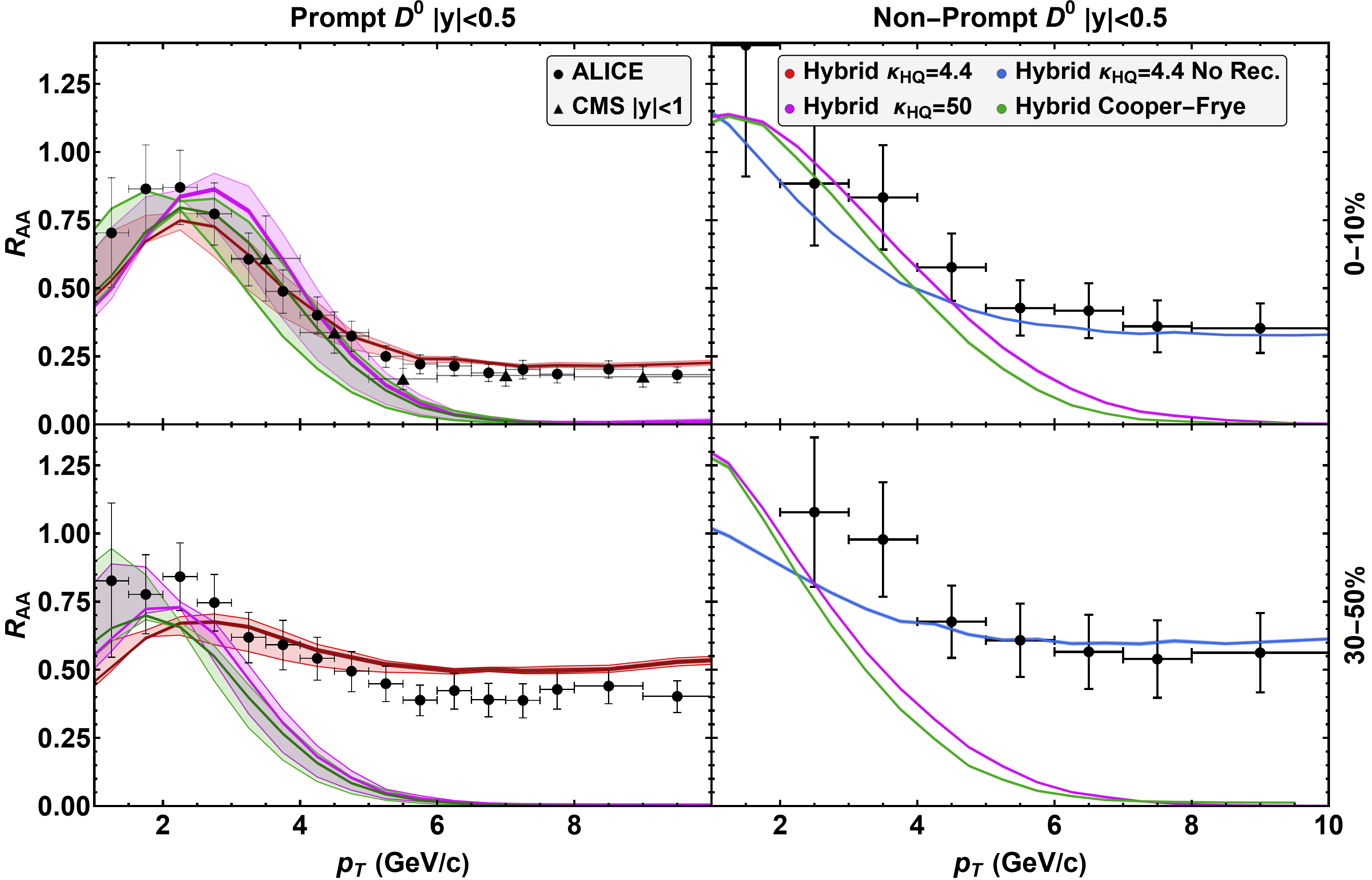}
    \caption{Results from Hybrid Model calculations with $\kappa_{\rm HQ}=4.4$ (red curves in the left panels, which are the same as the red curves in the left panels of Fig.~\ref{fig:hadraa}; blue curves in the right panels, which are the same as the blue curves in the right panels of Fig.~\ref{fig:hadraa}) and $\kappa_{\rm HQ}=50$ (pink curves) of the $R_{\rm AA}$ of prompt and non-prompt $D^0$ mesons, compared to ALICE \cite{ALICE:2020iug,ALICE:2023gjj} and CMS \cite{CMS:2020bnz,CMS:2022vfn} data, plotted as in Fig.~\ref{fig:hadraa}. The green curves show Hybrid Model results where the heavy quark momenta at the freezeout surface are sampled from the Cooper-Frye distribution, as described in the text.}
    \label{fig:d0raatherm}
\end{figure}

\begin{figure}
    \centering
    \includegraphics[width=\linewidth]{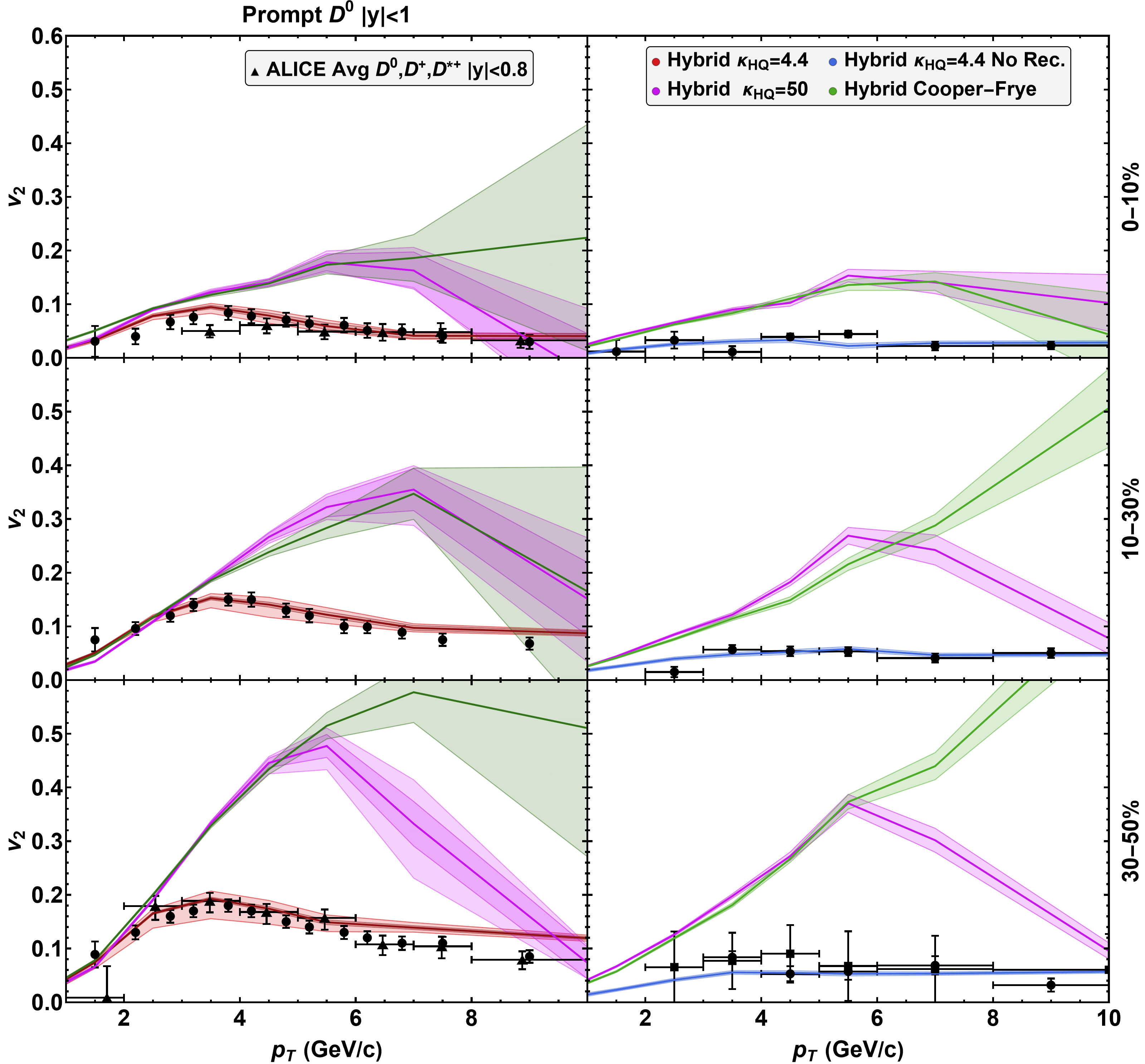}
    \caption{Results from Hybrid Model calculations with $\kappa_{\rm HQ}=4.4$ (red curves in the left panels, which are the same as the red curves in the left panels of Fig.~\ref{fig:hadv2} blue curves in the right panels, which are the same as the blue curves in the right panels of Fig.~\ref{fig:hadv2}) and $\kappa_{\rm HQ}=50$ (pink curves) of the $v_2$ of prompt and non-prompt $D^0$ mesons, compared to ALICE \cite{ALICE:2020iug,ALICE:2023gjj} and CMS \cite{CMS:2020bnz,CMS:2022vfn} data, plotted as in Fig.~\ref{fig:hadv2}. The green curves show Hybrid Model results where the heavy quark momenta at the freezeout surface are sampled from the Cooper-Frye distribution, as described in the text.}
    \label{fig:d0v2therm}
\end{figure}

There is a straightforward way in which we can elucidate the degree to which, and the range of $p_T$ in which, heavy quarks thermalize in our Hybrid Model calculations.  
To do so, we artificially crank the heavy quark drag coefficient up to $\kappa_{\rm HQ}=50$ and treat the $b$ and $c$ quarks as heavy (losing energy according to Eq.~\eqref{eq:heavyenergy-loss}) at all $p_T$,
rerun our calculations of $R_{\rm AA}$ and $v_2$, and compare the results we obtain to those that we have obtained above with $\kappa_{\rm HQ}=4.4$. 
If the charm (or bottom) quarks corresponding to prompt (or non-prompt) $D^0$ mesons with a given $p_T$ have thermalized in our Hybrid Model calculations with $\kappa_{\rm HQ}=4.4$, then cranking up $\kappa_{\rm HQ}$ (which makes thermalization happen more quickly) will result in no change to $R_{\rm AA}$ or $v_2$.  On the other hand, if the thermalization of charm or bottom quarks with $\kappa_{\rm HQ}=4.4$ is incomplete, cranking up $\kappa_{\rm HQ}$ by a large factor will accelerate their thermalization meaning that in the time they spend in the QGP they will thermalize more completely.  This means that the degree to which cranking $\kappa_{\rm HQ}$ up yields an increase in $v_2$ and a reduction in $R_{\rm AA}$ is a measure of the degree to which heavy quark thermalization is incomplete.
We perform this investigation in Figs.~\ref{fig:d0raatherm} and \ref{fig:d0v2therm}. By comparing the blue and pink curves in the right panels of Fig.~\ref{fig:d0v2therm}, we confirm that bottom quark thermalization is incomplete at all $p_T$, consistent with our previous conclusions.  
By comparing the red and pink curves in the left panels 
of Fig.~\ref{fig:d0v2therm}, we see that the charm quarks that hadronize as prompt $D^0$ mesons with $p_T\gtrsim 3$~GeV ($p_T\gtrsim 2.5$~GeV) have not fully thermalized in collisions with centralities up to 30\% (between 30 and 50\%).  And, 
for lower values of $p_T$ we see clear evidence for the thermalization of charm quarks. The agreement between  our Hybrid Model calculations and experimental measurements of $v_2$ at these lower values of $p_T$ indicates that charm quarks at these low $p_T$ thermalize in heavy ion collisions.
The comparison between the Hybrid Model calculations of $R_{\rm AA}$ in the pink and red curves in the left panels of Fig.~\ref{fig:d0raatherm} and the experimental data corroborates this conclusion for prompt $D^0$ mesons with $p_T\lesssim 3$~GeV.  

We can also ask out to what $p_T$ the charm quarks are, in fact, thermalized in the Hybrid Model calculation with $\kappa_{\rm HQ}=50$. We assess this via comparing the pink 
curves in Figs.~\ref{fig:d0raatherm} and \ref{fig:d0v2therm} to the green curves
obtained by taking the location of the charm quarks on the freezeout hypersurface from our $\kappa_{\rm HQ}=50$ Hybrid Model calculation but then choosing the prompt $D^0$ momenta via the Cooper-Frye freezeout procedure --- as if the charm quarks were perfectly thermalized at freezeout. From the comparison 
of pink to green in Fig.~\ref{fig:d0v2therm}, we see that
if $\kappa_{\rm HQ}$ were actually as large as 50, both charm and bottom quarks would thermalize for $p_T\lesssim 7$~GeV ($p_T\lesssim 6$~GeV) in heavy ion collisions with 
centralities of 0-30\% (30-50\%).  
The excellent agreement between the green curves
and the red curves and experimental data at $p_T\lesssim 2.5-3$~GeV provides further corroboration that charm quarks in this $p_T$ range equilibrate in heavy ion collisions.

\subsection{Choice of $\kappa_{\rm HQ}$}\label{sec:kappaval}

\begin{figure}
    \centering
    \includegraphics[width=\linewidth]{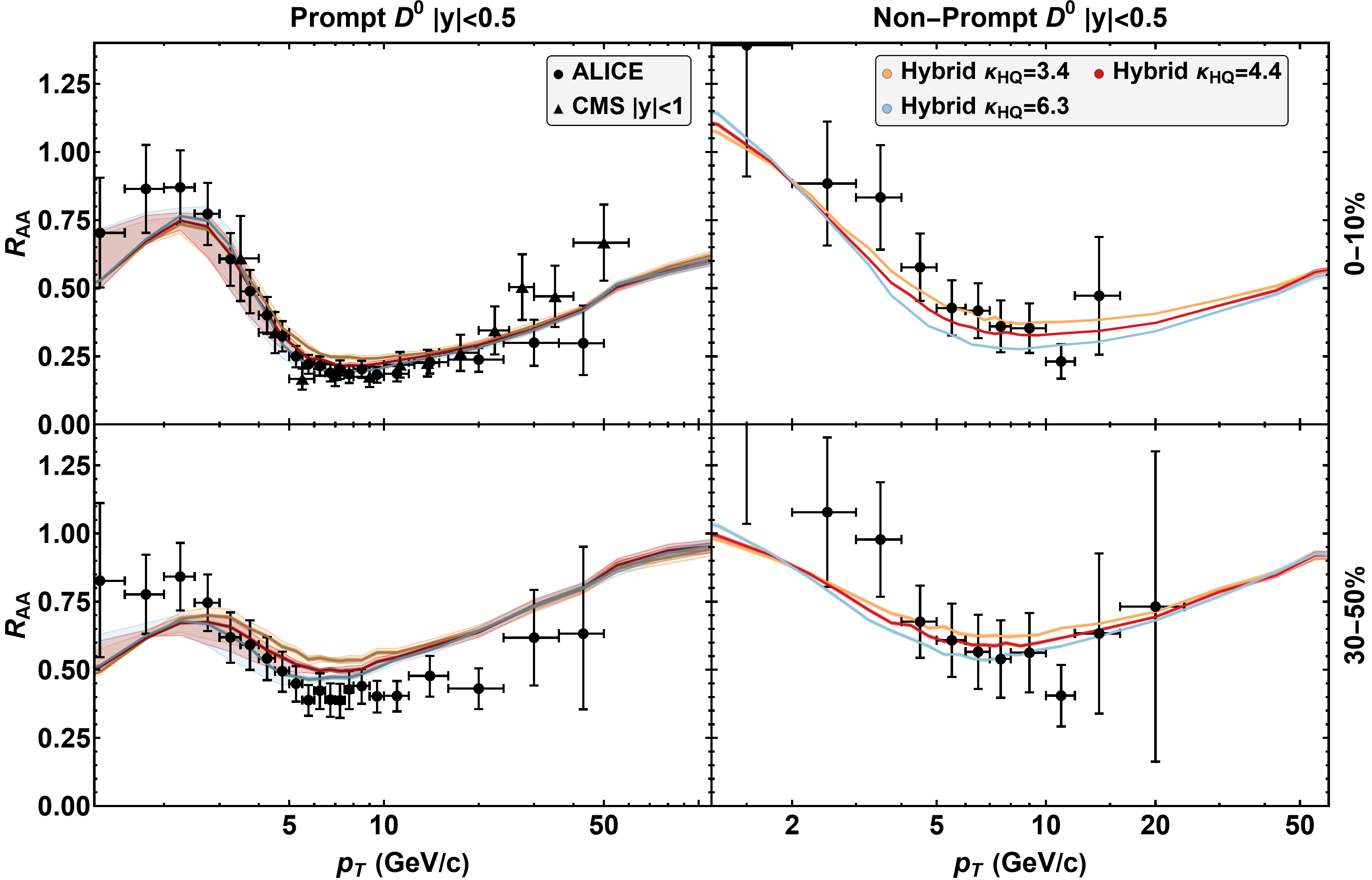}
    \caption{Results from Hybrid Model calculations with $\kappa_{\rm HQ}=3.4$, 4.4 (as in Fig.~\ref{fig:hadraa}) and 6.3 of the suppression $R_{\rm AA}$ of $D^0$ mesons compared to ALICE \cite{ALICE:2021rxa,ALICE:2021mgk} and CMS \cite{CMS:2017qjw} data, plotted as in Fig.~\ref{fig:hadraa}. Hybrid results and ALICE data use a rapidity cut of $|y|<0.5$, while the CMS data is for $|y|<1$.}
    \label{fig:d0raakappa}
\end{figure}

\begin{figure}
    \centering
    \includegraphics[width=\linewidth]{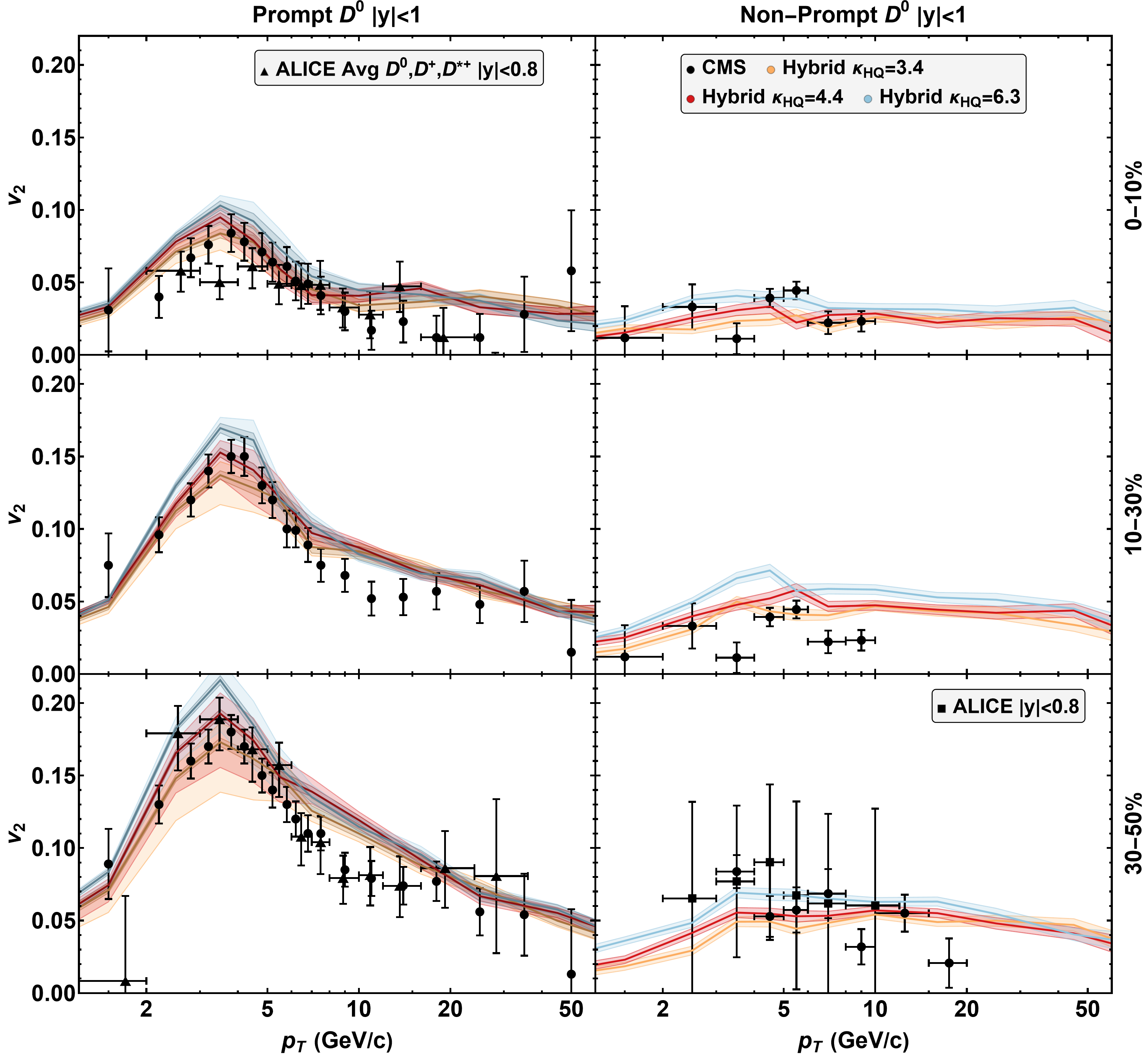}
    \caption{Results from Hybrid Model calculations with $\kappa_{\rm HQ}=3.4$, 4.4 (as in Fig.~\ref{fig:hadv2}) and 6.3 of the suppression $R_{\rm AA}$ of $D^0$ mesons compared to ALICE \cite{ALICE:2021rxa,ALICE:2021mgk} and CMS \cite{CMS:2017qjw} data, plotted as in Fig.~\ref{fig:hadv2}. Hybrid results and ALICE data use a rapidity cut of $|y|<0.5$, while the CMS data is for $|y|<1$.}
    \label{fig:d0v2kappa}
\end{figure}

Throughout Subsection~\ref{sec:mainresults} we have used the value $\kappa_{\rm HQ}=4.4$. In this Subsection, we motivate this choice by looking at the effect that varying $\kappa_{\rm HQ}$ from 3.4 to 6.3 has on the $R_{\rm AA}$ and $v_2$ results shown in Figs.~\ref{fig:hadraa} and~\ref{fig:hadv2}, respectively. In general terms,  increasing $\kappa_{\rm HQ}$ increases the amount of energy loss, and so one expects --- and  indeed observes in Figs.~\ref{fig:d0raakappa} and \ref{fig:d0v2kappa} --- a reduction of $R_{\rm AA}$ and an increase in $v_2$ with increasing $\kappa_{\rm HQ}$. 

The variation in $R_{\rm AA}$ and $v_2$ that results from varying $\kappa_{\rm HQ}$ from 3.4 to 6.3 is comparable to or smaller than the experimental uncertainties, meaning that the value of $\kappa_{\rm HQ}$ is not tightly constrained at present. However, we see from Fig.~\ref{fig:d0v2kappa} that the experimental measurements of $v_2$ are in tension with the choice $\kappa_{\rm HQ}=6.3$ in some ranges of $p_T$ and tend to favor $\kappa_{\rm HQ}=4.4$ or 3.4. In contrast, the experimental measurements of $R_{\rm AA}$ in the bottom-left panels of Fig.~\ref{fig:d0v2kappa}
are in tension with the choice $\kappa_{\rm HQ}=3.4$ in some ranges of $p_T$ and tend to favor $\kappa_{\rm HQ}=6.3$ and 4.4. From these comparisons, it seems reasonable to estimate that the best choice of $\kappa_{\rm HQ}$ lies somewhere in the range $3<\kappa_{\rm HQ}<6$, but we have not attempted a formal Bayesian quantification of the uncertainty. 
We defer an analysis of the constraints on $\kappa_{\rm HQ}$ and $\kappa_{\rm sc}$ arising from experimental measurements of heavy hadron, light hadron and jet observables using Bayesian inference techniques to future work.
For the present paper, we have chosen to use $\kappa_{\rm HQ}=4.4$, noting that smaller values worsen the description of $R_{\rm AA}$ for prompt $D_0$ mesons in the 30-50\% centrality class, see Fig.~\ref{fig:d0raakappa}, whereas larger values worsen the description of the $v_2$ of both prompt and non-prompt $D^0$ mesons, see Fig.~\ref{fig:d0v2kappa}.

We can also ask whether $\kappa_{\rm HQ}=4.4$ seems reasonable based upon what we know from ${\cal N}=4$ supersymmetric Yang-Mills (SYM) theory.  To set a context for this, let us recall that the value $\kappa_{\rm sc}=0.404$ that was obtained in Ref.~\cite{Casalderrey-Solana:2018wrw} via fitting Hybrid Model calculations of the suppression of light hadrons and inclusive jets to data corresponds to a light quark stopping length $x_{\rm stop} \propto 1/\kappa_{\rm sc}$ (see Eqs.~\eqref{eq:elossrate} and \eqref{eq:xstop}) that is 3.9 times longer in the QGP of QCD than in the strongly coupled plasma of ${\cal N}=4$ SYM theory with 't Hooft coupling $\lambda=12$. 
This seems reasonable given the larger number of degrees of freedom in the ${\cal N}=4$ SYM plasma.
The stopping length for a heavy quark in ${\cal N}=4$ SYM theory is $\propto 1/\eta_D \propto 1/\kappa_{\rm HQ}$, and $\kappa_{\rm HQ}^{{\cal N}=4}=\pi\sqrt{\lambda}$. This means that the choice $\kappa_{\rm HQ}=4.4$ corresponds to a heavy quark stopping length that is 2.5 times longer in the QGP of QCD than in the strongly coupled plasma of ${\cal N}=4$ SYM theory with $\lambda=12$, which again seems reasonable. Of course, any value in the range $3<\kappa_{\rm HQ}<6$ would be similarly reasonable by this argument.

At vanishing velocity we can use the Einstein relation to relate a value of $\kappa_{HQ}=4.4$ to a spatial diffusion constant of $2\pi T D = 4\pi/\kappa_{\rm HQ} =  2.9$, which we can compare to other determinations of $D$. This value is in good agreement with previous phenomenological extractions of constraints on $D$ from experimental data~\cite{Liu:2016ysz,Xu:2017obm,Ke:2018tsh,Cao:2018ews,Tang:2023tkm,ALICE:2021rxa,Sambataro:2025obe}.
And, as in this prior work,
$2\pi T D \sim 3$ is in some tension with lattice QCD calculations which find values of $2\pi T D$ between 1 and 2 for temperatures in the range 175-220~MeV~\cite{Altenkort:2023eav,HotQCD:2025fbd}.
We note that all phenomenological extractions, ours included, employ measurements of heavy flavor hadrons with nonzero velocity whereas the lattice calculations are performed for heavy quarks with zero velocity.
This suggests that the tension 
between phenomenological extractions of $D$ and the value obtained from lattice calculations may reflect 
the non-Gaussianity in the fluctuations in the momentum transferred between a heavy quark and the medium 
that has been shown to be significant in strongly coupled ${\cal N}=4$ SYM theory for heavy quarks at nonzero velocity~\cite{Rajagopal:2025ukd}
and that modifies the Einstein relation for heavy quarks at nonzero velocity in any quantum field theory~\cite{Rajagopal:2025rxr}.
Such non-Gaussian fluctuations have not yet been incorporated in any of the extractions of $D$ from experimental data.

We have now completed our principal goals in this work: we have introduced heavy quarks into the Hybrid Model with a new composite expression describing how they lose energy and a conventional approach to describing their momentum diffusion; we have added the possibility that heavy quarks may hadronize either via fragmentation or via LCN recombination with a light antiquark or diquark from the flowing droplet of QGP; and, we have shown that with a reasonable choice for the value of the new model parameter that governs heavy quark energy loss we obtain a good description of extant experimental measurements of the suppression, partial thermalization, and elliptic flow of charmed and bottom hadrons (in the latter case, with non-prompt $D^0$ mesons as a proxy).

Now with a model framework in which we can calculate heavy flavor and jet dynamics in place, there are many additional observables that we can now compute and describe within the Hybrid Model.  In the next Subsections, we investigate three.

\subsection{Non-Prompt $J/\psi$ suppression}

\begin{figure}
    \centering
\includegraphics[width=0.6\linewidth]{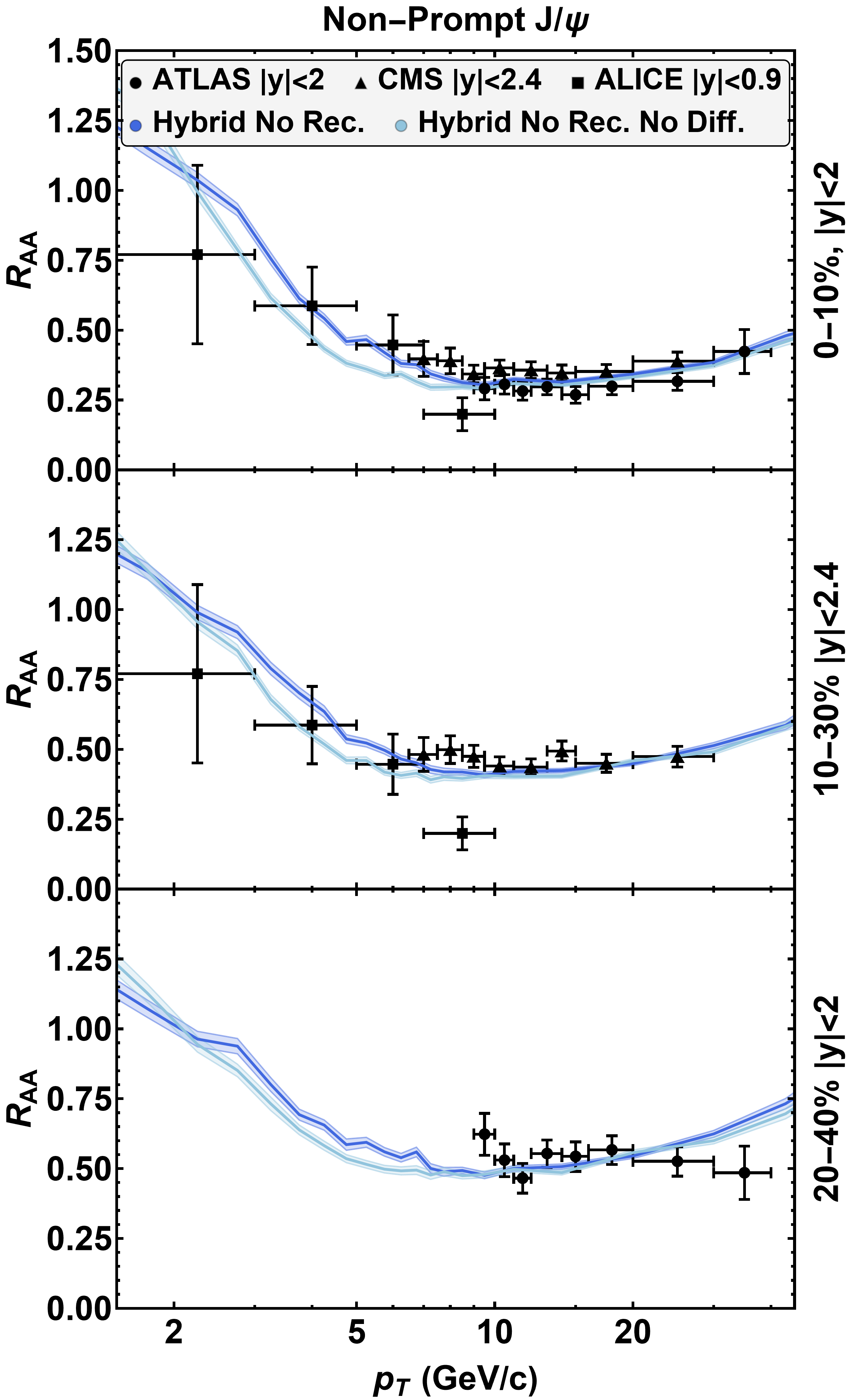}
    \caption{Results of Hybrid Model calculations with $\kappa_{\rm HQ}=4.4$ of the suppression $R_{AA}$ of non-prompt $J/\psi$ mesons in PbPb collisions at $\sqrt{s_{\rm NN}}=5.02$~TeV with (dark blue) and without (light blue) momentum diffusion compared to ATLAS \cite{ATLAS:2018hqe}, CMS \cite{CMS:2017uuv} and ALICE \cite{ALICE:2023hou} data. The centrality and rapidity cuts used in our Hybrid model calculations are labeled on the right-hand side of each panel, and all data reported in each panel is at the same centrality. The rapidity cuts used by each experiment are listed in the legend at the top. For each experimental data point, we have added reported uncertainties in quadrature.}
    \label{fig:npjpsi}
\end{figure}

To this point, we have relied upon non-prompt $D$ mesons as proxies for $b$ quarks.  $b$ quark decays can also yield 
non-prompt $J/\psi$ mesons, 
however, making them comparably good proxies for $b$ quarks. The $R_{\rm AA}$ of non-prompt $J/\psi$ mesons
has been measured in PbPb collisions at the LHC by ATLAS, ALICE and CMS~\cite{ATLAS:2018hqe,ALICE:2023hou,CMS:2017uuv}. This observable is sensitive to the energy loss and momentum diffusion of $b$ quarks, while being insensitive to recombination.
And, the uncertainties in the measurements of $R_{\rm AA}$ for non-prompt $J/\psi$ mesons~\cite{ATLAS:2018hqe,ALICE:2023hou,CMS:2017uuv} are smaller than the uncertainties in the measurements of $R_{\rm AA}$ for non-prompt $D$ mesons~\cite{ALICE:2021mgk} that we have compared our calculations to in previous subsections. We have fixed the 
value $\kappa_{\rm HQ}=4.4$ based upon our comparisons to prompt $D$ meson (which is to say charm quark) suppression and flow, which
makes it interesting to compare our calculations of $R_{\rm AA}$ for non-prompt $J/\psi$ mesons to the  data, as we do in 
Fig.~\ref{fig:npjpsi}, as doing so provides a somewhat independent test of our model.
We see in Fig.~\ref{fig:npjpsi} that our calculations with $\kappa_{\rm HQ}=4.4$ are in good agreement with the experimental measurements,
both with and without momentum diffusion.  That said, including momentum diffusion does serve to improve the agreement with data somewhat in 
collisions with $0-10\%$ centrality. 
As the uncertainties in these experimental measurements improve with the higher statistics data sets to come from LHC Run 4 and in Run 5
with the ALICE 3 detector, and with the addition of high-statistics measurements of the elliptic flow of non-prompt $J/\psi$ mesons,
$b$ quark measurements should
yield as tight constraints on heavy quark energy loss and momentum diffusion and hence the value of $\kappa_{\rm HQ}$ as $c$ quark measurements do, while being insensitive to recombination effects.

\subsection{$b$-jet RAA}
\label{sec:jetRAA}

\begin{figure}[t]
    \centering
    \includegraphics[width=\linewidth]{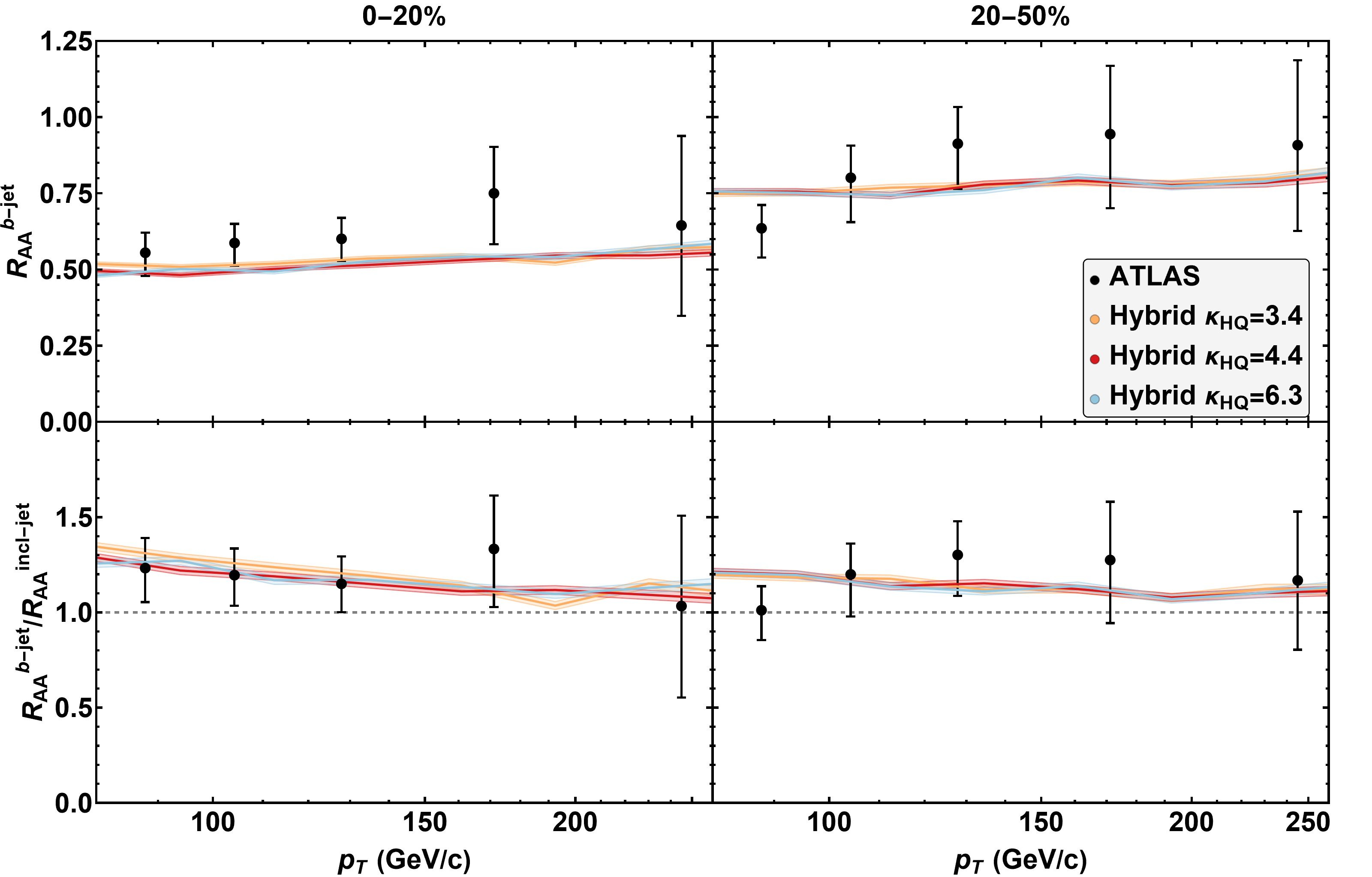}
    \caption{Results of Hybrid Model calculations with three  different values of $\kappa_{\rm HQ}$ for the suppression $R_{\rm AA}$ of $b$-jets (upper panels) and for the ratio of the suppression of $b$-jets to the suppression of inclusive jets (lower panels) compared to ATLAS measurements of anti-$k_t$ jets with $R=0.2$ in heavy ion collisions with $\sqrt{s_{\rm NN}}=5.02$~TeV with 0-20\% centrality and $|y|<2.1$~\cite{ATLAS:2022agz}. For each experimental data point, we have added the statistical and systematic uncertainties in quadrature.}
    \label{fig:bjets}
\end{figure}

Having discussed the two most basic and fundamental heavy quark observables in Subsection~\ref{sec:mainresults}, namely single hadron $R_{\rm AA}$ and $v_2$, we want to take advantage of the capabilities of the Hybrid Model and extend our predictions to those involving high-$p_T$ jets that contain a $b$ quark, tagged via the displaced vertex that characterizes the decay of a bottom meson or baryon. In Fig.~\ref{fig:bjets}, we look at the suppression of $b$-tagged jets measured by ATLAS~\cite{ATLAS:2022agz}, and how their suppression compares to the suppression of inclusive jets. We see that the Hybrid Model results are completely insensitive to the value of $\kappa_{\rm HQ}$. This is to be expected
because, according to the composite description of energy loss in Eq.~\eqref{eq:composite}, most of the $b$ quarks in jets
with the large values of $p_T$ in
Fig.~\ref{fig:bjets} lose energy as if they were light quarks, with no dependence on $\kappa_{\rm HQ}$.

We see that we seem to be slightly over-quenching $b$-jets in central collisions, but in a way that is consistent with the over-quenching of inclusive jets, giving very reasonable agreement in the double ratio of their $R_{\rm AA}$. 
One of the reasons why these Hybrid Model calculations slightly over-estimate the quenching of jets
is that we are running the Hybrid Model without including the wakes of jets, introduced in Ref.~\cite{Casalderrey-Solana:2016jvj}. Jet wakes make no contribution to the heavy hadron observables that are our sole focus elsewhere in this paper.  Jet wakes do contribute to reconstructed jets, meaning that neglecting the wakes over-estimates their quenching. This is a small effect for anti-$k_t$ jets with $R=0.2$, as for jets that are this narrow most of the energy in the jet wakes will be outside of the jet cone.

Most importantly, the agreement between the results of our Hybrid Model calculations and ATLAS data for the ratio of $b$ jets to inclusive jets serves as a confirmation that the Hybrid Model treatment of the energy loss of quark-initiated jets relative to that of gluon-initiated jets is in agreement with data.  The results in Fig.~\ref{fig:bjets}, together with our heavy hadron results from Subsection \ref{sec:mainresults} and the many results from a decade of Hybrid Model calculations of jet and light hadron observables establish that in the Hybrid Model we now have a seamless framework with which to obtain a quantitative description of heavy flavor and jet probes of QGP, the dynamical response of the droplets of QGP to these probes, and their observable consequences.   

\subsection{$\Lambda_c/D^0$, $D^+/D^0$ and $D^+_s/D^0$ 
Ratios}

\begin{figure}
    \centering
    \includegraphics[width=\linewidth]{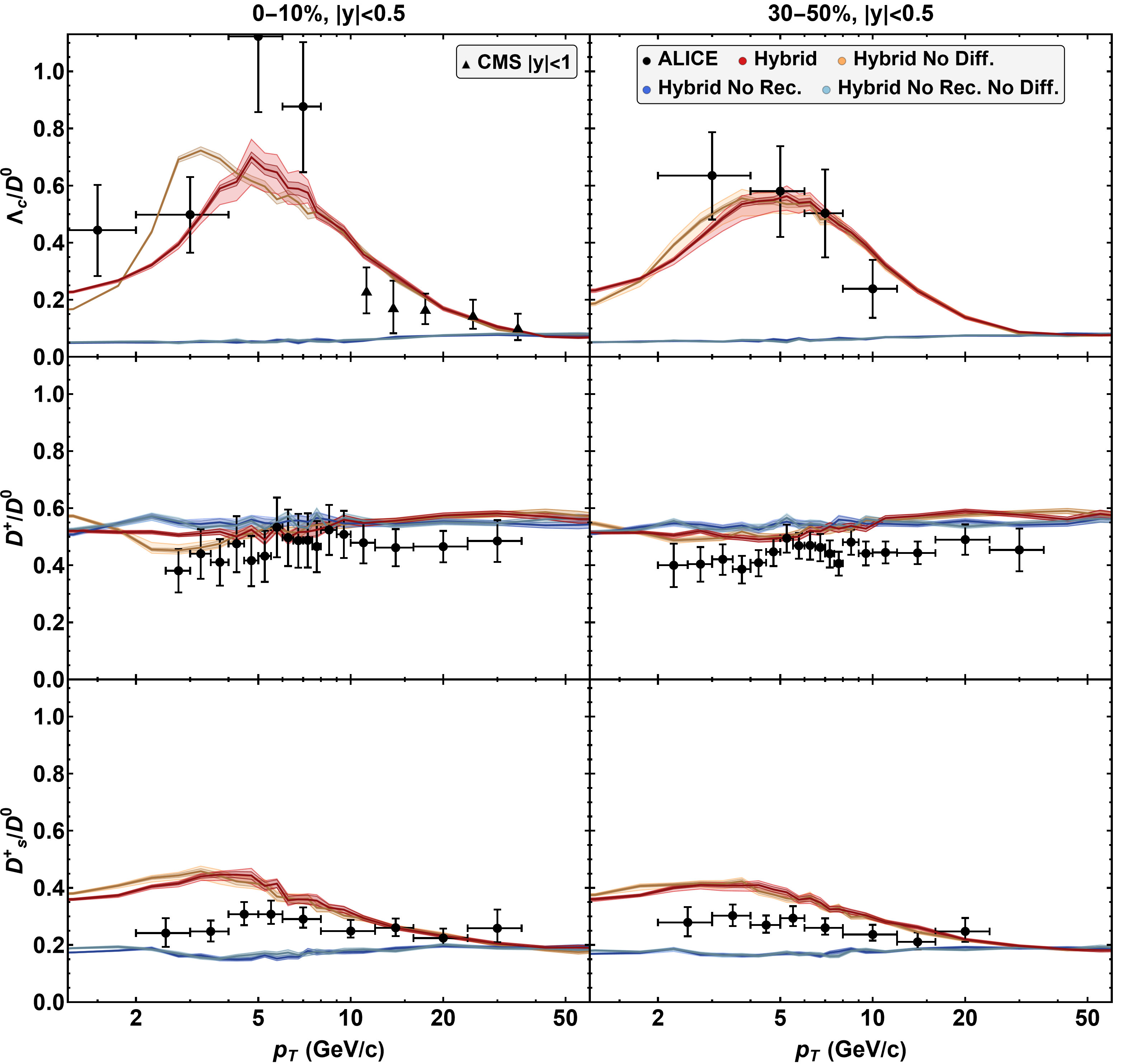}
    \caption{Results of Hybrid Model calculations with $\kappa_{\rm HQ}=4.4$ of the $\Lambda_c/D^0$, $D^+/D^0$ and $D^+_s/D^0$ ratios as a function of $p_T$ in heavy ion collisions with $\sqrt{s_{\rm NN}}=5.02$~TeV with 0-10\% (left panels) and 30-50\% (right panels) centrality,  compared to ALICE \cite{ALICE:2021bib,ALICE:2021rxa,ALICE:2021kfc} and CMS \cite{CMS:2023frs} data. Hybrid results and ALICE data use a rapidity cut of $|y|<0.5$, while the CMS data is for $|y|<1$. For each experimental data point, we have added the statistical and systematic uncertainties in quadrature.
    Each panel shows four Hybrid Model calculations, with colors as in Figs.~\ref{fig:hadraa} and \ref{fig:hadv2}: the red curves correspond to the full Hybrid Model calculation; the blue curves show that it is impossible to describe the $\Lambda_c/D^0$ data without including hadronization by recombination.}
    \label{fig:hadratios}
\end{figure}

Another class of observables that we can now compute with the Hybrid Model is 
heavy baryon-to-meson and heavy meson-to-meson ratios, as a function of $p_T$. By comparing calculations of these ratios to experimental data we are primarily looking at the hadrochemistry encapsulated by the LCN model of recombination that we employ, although effects originating from the transport of charm quarks do also play a role.
In Fig.~\ref{fig:hadratios}, we see that, as is familiar from previous work~\cite{Beraudo:2022dpz}, the LCN model for recombination over-predicts the $D^+_s/D^0$ ratio at low $p_T$ because to date it does not include the possibility that clusters formed from a charm quark and a strange antiquark may decay to a $D$ meson and a kaon.
In Fig.~\ref{fig:hadratios} we also see that we 
obtain a rather good description of the $\Lambda_c/D^0$ and 
$D^+/D^0$ ratios --- and we see that in the former case this agreement depends crucially on including hadronization via recombination.  In the LCN recombination model, a charm quark can hadronize together with a light diquark from the QGP. Here, we see that including this effect is crucial to describing heavy baryon-to-meson ratios.
We note that the behavior we see is fully consistent with previous results obtained using LCN recombination with different models of charm quark transport~\cite{Beraudo:2022dpz}.  The novelty here is not the description of the $\Lambda_c/D^0$ ratio at intermediate $p_T$, it is the fact that we obtain this description in the Hybrid Model in which we can at the same time compute high-$p_T$ light and heavy hadron and jet observables.

\subsection{Charmed baryon flow}

\begin{figure}
    \centering
    \includegraphics[width=\linewidth]{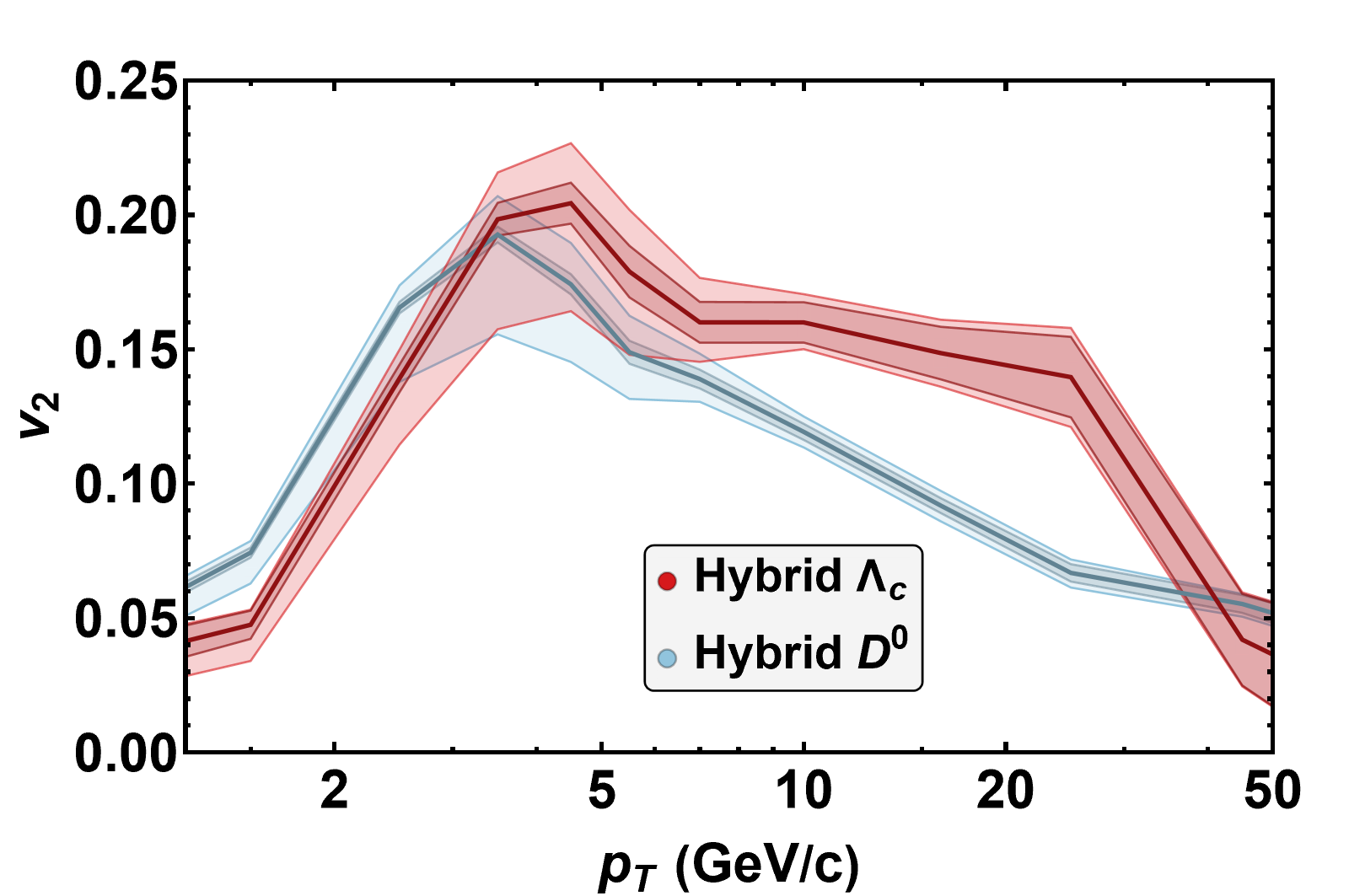}
    \caption{Results of Hybrid Model calculations $\kappa_{\rm HQ}=4.4$ of the ellliptic flow $v_2$ of $\Lambda_c$ baryons in heavy ion collisions with $\sqrt{s_{\rm NN}}=5.02$~TeV with 30-50\% centrality and $|y|<1.0$. For reference, we also show Hybrid Model calculations of $v_2$ of prompt $D^0$ mesons: the blue band here is the same as the red band in the bottom-left panel of Fig.~\ref{fig:hadv2}. The $\Lambda_c$ elliptic flow that we obtain is in excellent agreement with preliminary data from ALICE~\cite{lambdaflow}.}
    \label{fig:lambdaflow}
\end{figure}

Motivated by preliminary data from a new measurement by ALICE~\cite{lambdaflow}, the
final observable that we calculate is the elliptic flow $v_2$
for $\Lambda_c$ baryons. We show results from our Hybrid Model calculations of these observables in Fig.~\ref{fig:lambdaflow}, comparing them there to our calculations of the elliptic flow of $D^0$ mesons from Fig.~\ref{fig:hadv2}.  As is familiar from comparisons of long standing between the elliptic flow of light baryons and mesons, and as has often been attributed in that case to the fact that baryons contain three constituent quarks, the $v_2(p_T)$ for $\Lambda_c$ baryons is shifted to the right and peaks somewhat higher than that for $D^0$ mesons.  
Furthermore, the results in Fig.~\ref{fig:lambdaflow} are in qualitative and quantitative agreement with preliminary results from the new ALICE measurement~\cite{lambdaflow}. 
Within our Hybrid Model, and in particular via the inclusion of LCN hadronization by recombination, the crossing of the $D^0$ and $\Lambda_c$ $v_2$ curves as a function of increasing $p_T$ (with charmed baryon $v_2$ going from below charmed meson $v_2$ to above it)
has to be attributed to the larger flow of the thermal diquarks involved in the recombination process. 
It is notable that, even after accounting for our systematic theoretical uncertainties, the $D^0$ and $\Lambda_c$ curves remain well separated up to $p_T\sim 20$ GeV/c, suggesting the presence of important effects arising from in-medium hadronization even at these values of transverse momentum.
It will be important to see whether the final ALICE data supports this picture, as the preliminary data does~\cite{lambdaflow}. 
This looks consistent with the relative contribution of recombination and fragmentation processes to charmed hadron production that we see in Fig.~\ref{fig:hadform}.

\section{Taking Stock and Looking Ahead}
\label{sec:Ahead}

Heavy quarks offer an invaluable hard probe of the droplets of quark gluon plasma formed in heavy ion collisions at the LHC and RHIC. Given their large mass,  they are predominantly produced in hard scattering processes at the earliest moment of a collision and given their rarity they almost never annihilate with a heavy antiquark subsequently. This means that they experience, and probe, the entire history of the expanding, cooling, droplet of QGP from hydrodynamization through hadronization. Quantitative measurements of heavy quark final state observables therefore give us access to information about the transport properties of QGP
as well as about medium modifications of hadronization. Before this work, the Hybrid 
Model of jet quenching has not included any implementation of the heavy-quark sector, which has made it impossible to confront its predictions with measurements of heavy quark and jet observables together, in a unified fashion.

We have introduced heavy quarks into the Hybrid Model for the first time, having made this possible by introducing a new composite expression describing how they lose energy in strongly coupled QGP.
Our formulation of heavy quark energy loss is informed by two
long-established holographic results from strongly coupled ${\cal N}=4$ SYM theory, one of which provides the 
rate of energy loss for a massless quark, which should 
be a good description of how an ultrarelativistic heavy quark loses energy, and the other of which provides the drag force on an infinitely massive quark, which should describe the rate of energy loss of a heavy quark slowing down and coming to rest well.
We have introduced a new composite energy loss formula that incorporates the two limits in a way that ensures that the energy of the heavy quark, and its first derivative, change continuously as a function of time or distance. 
We have 
implemented this composite energy loss formula into the Hybrid Model together with a conventional Gaussian treatment of momentum diffusion, ensuring that diffusion and drag are related by the Einstein relation so as to ensure that (given sufficient time) the heavy quarks will end up at rest in the local fluid rest frame, diffusing with a Boltzmann distribution for their momenta. 
With the composite expression for energy loss and 
momentum diffusion implemented, we can describe
a heavy quark that is initially ultrarelativistic and losing energy as if it were a light quark, subsequently loses energy as heavy quarks do, gets dragged to rest in, and ends up diffusing in, the flowing fluid.
With the Hybrid Model augmented in this way so as to describe heavy quark transport, in order to describe the phenomenology of heavy mesons and baryons at low and intermediate $p_T$ we have incorporated hadronization of heavy quarks via recombination using the Local Color Neutralization model. 
With these developments in place, in the Hybrid Model we now have a common framework for seamless modeling of jets, jet substructure, jets containing a heavy quark, light hadrons, and heavy hadrons at high and low $p_T$ in heavy ion collisions.

There are of course many improvements to our treatment of heavy quarks that can be pursued in future work.  We have neglected $T/M$ corrections to both the draft coefficient and to the Einstein relation that relates it to the longitudinal diffusion coefficient. Furthermore, although our implementation of momentum diffusion in terms of Gaussian momentum fluctuations is conventional, it is 
not adequate because the fluctuations in the momentum transferred between a relativistic heavy quark and the strongly coupled plasma through which it propagates are non-Gaussian~\cite{Rajagopal:2025ukd,Rajagopal:2025rxr}. Implementing $T/M$ corrections as well as  non-Gaussian fluctuations in the Hybrid Model remains for the future. The present uncertainties in the experimental measurements of heavy flavor observables are not yet so small as to mandate these improvements.  The LCN treatment of recombination that we have employed can also be improved. In particular, we have found that at mid-centralities and intermediate momenta $p_T\sim 10$~GeV, our model with recombination turned off does somewhat better when compared to data, which motivates investigating possible improvements to the straightforward procedure for switching between hadronization via recombination and hadronization via fragmentation that we have employed. 

In the present work, we have confronted the newly augmented Hybrid Model with experimental measurements from ALICE, ATLAS and CMS of the 
suppression $R_{\rm AA}$ and elliptical anisotropy $v_2$
of prompt $D$-mesons, $B$-mesons (via non-prompt $D^0$ mesons) and $\Lambda_c$ baryons as well as heavy baryon-to-meson ratios and $R_{\rm AA}$
of $b$-jets.  We find that when we include both momentum diffusion and LCN recombination we obtain broad, and good, agreement with extant experimental measurements of these observables across the whole available range of transverse momenta and collision centrality upon choosing a reasonable value of $\kappa_{\rm HQ}$, the new model parameter that governs heavy quark energy loss.  
Although the present uncertainties in the experimental measurements are too large to yield a tight constraint on the value of $\kappa_{\rm HQ}$, it is pleasing to find that 
with $\kappa_{\rm HQ}=4.4$, which corresponds to a heavy quark stopping length in QGP that is 2.5 times longer than in the strongly coupled ${\cal N}=4$ SYM plasma at the same temperature, the
measurements of the suppression, partial thermalization, and elliptic flow of charmed and
bottom mesons and charmed baryons, the suppression of $b$-jets, and heavy baryon-to-meson ratios, are all described well.
By comparing the results of our calculations to results obtained upon artificially increasing $\kappa_{\rm HQ}$ by more than a factor of ten so as to accelerate the thermalization of heavy quarks, we have also confirmed that 
prompt $D^0$ mesons with $p_T\lesssim 3$~GeV have thermalized in the droplets of QGP produced in heavy ion collisions at the LHC, whereas $B$ mesons have not.

This work provides a seamless model framework in which it is possible to calculate heavy flavor and jet dynamics, together. With this in place, there are many additional observables that can be computed and described within the Hybrid Model. Obvious examples include further studies of jets, and the substructure of jets, that contain heavy quarks. One example is the angular decorrelation between the $D^0$-direction and the jet axis for events with $D^0$ mesons in jets, which has been measured by CMS~\cite{CMS:2019jis}.
It would also be very interesting to investigate photon-$D$, $Z$-$D$ and $D$-$\bar D$ correlations, as their angular decorrelation should provide access to effects of momentum diffusion and hadronization.  
Not that further motivation is needed, but a big reason (in addition to the fact that we now have the framework within which to pursue them)  why such studies would be timely is that these and many other heavy flavor observables will be measured soon by sPHENIX, meaning that we will be able to compare what we learn from heavy ion collisions at RHIC to what we are learning from heavy ion collisions at the LHC.

A further direction that is very much worth pursuing is a formal Bayesian uncertainty quantification of the parameters in the Hybrid Model including $\kappa_{\rm sc}$ and $\kappa_{\rm HQ}$, which govern the energy loss of light partons and heavy quarks, as well as $L_{\rm res}$, the resolution length of QGP, and the two parameters needed to describe large-angle elastic (Moli\`ere) scattering of partons from the jet shower off quasiparticles from the QGP that takes full advantage of the extant experimental measurements of the by now numerous and varied suite of jet, jet substructure and heavy flavor observables.  As heavy flavor measurements from sPHENIX become available and as the uncertainties on the measurements from the LHC experiments decrease with increasing statistics, the constraints on $\kappa_{\rm HQ}$ obtained via such a Bayesian uncertainty quantification will tighten.

\acknowledgments

We are grateful to Carlos Hoyos, Gian Michele Innocenti, Yen-Jie Lee and Bruno Scheihing-Hitschfeld for helpful conversations. KR acknowledges the hospitality of the CERN Theory Department and the Aspen Center for Physics.
Research supported in part by the U.S.~Department of Energy, Office of Science, Office of Nuclear Physics under grant Contract Number DE-SC0011090.
Research of DP supported in part by the European Union's Horizon 2020 research and innovation program under the Marie Sk\l odowska-Curie grant agreement No 101155036 (AntScat), by the European Research Council project ERC-2018-ADG-835105 YoctoLHC, by the Spanish Research State Agency under project 
PID2020-119632GB-I00, by Xunta de Galicia (CIGUS Network of Research Centres) and the European Union, by Unidad de Excelencia Mar\'ia de Maetzu under project CEX2023-001318-M, and by the Ram\'on y
Cajal fellowship RYC2023-044989-I.
A.B. acknowledges financial support by MUR within
the Prin 2022sm5yas project.
This work made use of  
SubMIT~\cite{Bendavid:2025arv} in the MIT Department of Physics.

\appendix

\section{Details of the implementation of energy loss}
\label{sec:AppendixA}

The formulae in the 
expression~\eqref{eq:composite} for energy loss are straightforward to apply to on-shell particles, but we need to specify how we apply them to the off-shell particles found in a parton shower.  This was done for massless partons in the Hybrid Model in Refs.~\cite{Casalderrey-Solana:2014bpa,Casalderrey-Solana:2015vaa}.
In the Hybrid Model, on-shell and off-shell massless partons 
lose energy in the local fluid rest frame during each time step $\Delta t$ according to Eq.~\eqref{eq:elossrate}, which is to say
\begin{equation}
    \Delta E=-\frac{4 x^2 E_{\rm in}}{\pi x_{\rm stop}^2\sqrt{x_{\rm stop}^2-x^2}}\Delta t \ .
    \label{eq:Appendix-light-energy-loss}
\end{equation}
In the case of a massless parton, the three-momentum then decreases as
\begin{equation}
    \Delta\vec{p}=\frac{\Delta E}{E}\vec{p} \ .
    \label{eq:Appendix-light-momentum-loss}
\end{equation}
This keeps an on-shell  massless parton on-shell, rescales its momentum by the appropriate amount, and does not deflect the direction of the parton --- all as expected. The virtuality of an off-shell massless parton is reduced by a factor  $(E-\Delta E)/E$, and off-shell partons are also not deflected. 
Since the effect at each timestep is simply a rescaling of the four-momentum, the direction of the parton in the lab frame will also not be changed by energy loss. This is all as one expects. 

In the present work, however, we wish to
employ Eq.~\eqref{eq:elossrate} to describe the energy loss of a high energy parton that has a nonzero mass, which is to say we wish to describe the energy loss of a high-energy heavy quark that is behaving as a light quark. 
In this case, 
there is a remaining subtlety that prompts us to modify how we decrease the momentum of a high-energy quark with nonzero mass that loses energy according to Eq.~\eqref{eq:Appendix-light-energy-loss}. 
We want to ensure that as the heavy quark loses energy in this fashion, its virtuality does not increase.
We do so by modifying the expression~\eqref{eq:Appendix-light-momentum-loss} for how the heavy quark that is being treated as a light quark loses three-momentum to
\begin{equation}
    \Delta\vec{p}=\frac{E\Delta E}{\vec{p}^2}\vec{p}\ .
\end{equation}
This is equivalent to Eq.~\eqref{eq:Appendix-light-momentum-loss} for massless quarks, but for off-shell quarks with a nonzero mass it ensures
that
the virtuality of the quark always decreases.

For heavy quarks that lose energy according to Eq.~\eqref{eq:heavyenergy-loss}, a simple rescaling of the 
four-momentum is no longer an appropriate choice. Additionally, a rescaling of the three-momentum of a heavy quark in the local fluid rest frame will in general look like a deflection in the lab frame.  This contributes to low-energy heavy quarks being dragged along by a moving fluid, behavior that is expected.
We have chosen to implement Eq.~\eqref{eq:heavyenergy-loss} for heavy quarks as follows. In the local fluid rest frame, energy is lost according to
\begin{equation}
    \Delta E=-\eta_D\,\frac{\vec{p}^2}{E}\,\Delta t \ .
\end{equation}
For an on-shell quark, this is equivalent to Eq.~\eqref{eq:heavyenergy-loss}. For the three-momenta, we choose
\begin{equation}
    \Delta\vec{p}=-\eta_D\,\vec{p}\,\Delta t \ ,
\end{equation}
which is again the correct expression for an on-shell quark.
As with light partons in the Hybrid Model, 
we include heavy quark energy loss and diffusion only as long as the temperature of the QGP at the location of the heavy quark is above $145$ MeV.

\bibliography{arxivv1.bib}
\bibliographystyle{JHEP}

\end{document}